\documentclass[prb,superscriptaddress,showpacs,floatfix,tightenlines,10pt,twocolumn]{revtex4}
\usepackage{amssymb}
\usepackage{amsmath}
\usepackage{graphicx}

\begin{document}

\title{Phase diagram of insulating crystal and quantum Hall states in
ABC-stacked trilayer graphene }
\author{R. C\^{o}t\'{e} }
\affiliation{D\'{e}partement de physique, Universit\'{e} de Sherbrooke, Sherbrooke, Qu%
\'{e}bec, J1K 2R1, Canada}
\author{Maxime Rondeau }
\affiliation{D\'{e}partement de physique, Universit\'{e} de Sherbrooke, Sherbrooke, Qu%
\'{e}bec, J1K 2R1, Canada}
\author{Anne-Marie Gagnon}
\affiliation{D\'{e}partement de physique, Universit\'{e} de Sherbrooke, Sherbrooke, Qu%
\'{e}bec, J1K 2R1, Canada}
\author{Yafis Barlas}
\affiliation{Department of Physics and Astronomy, University of California, Riverside, CA
92521}

\begin{abstract}
In the presence of a perpendicular magnetic field, ABC-stacked trilayer
graphene's chiral band structure supports a 12-fold degenerate $N=0$ Landau
level (LL). Along with the valley and spin degrees of freedom, the zeroth LL
contains additional quantum numbers associated with the LL orbital index $%
n=0,1,2$. Remote inter-layer hopping terms and external potential difference 
$\Delta _{B}$ between the layers lead to LL splitting by introducing a gap $%
\Delta _{LL}$ between the degenerate zero-energy triplet LL orbitals.
Assuming that the spin and valley degrees of freedom are frozen, we study
the phase diagram of this system resulting from competition of the single
particle LL splitting and Coulomb interactions within the Hartree-Fock
approximation at integer filling factors. Above a critical value $\Delta
_{LL}^{c}$ of the external potential difference i,e, for $\left\vert \Delta
_{LL}\right\vert >\Delta _{LL}^{c}$, the ground state is a uniform quantum
Hall state where the electrons occupy the lowest unoccupied LL orbital
index. For $\left\vert \Delta _{LL}\right\vert <\Delta _{LL}^{c}$ (which
corresponds to large positive or negative values of $\Delta _{B}$) the
uniform QH state is unstable to the formation of a crystal state at \emph{%
integer} filling factors. This phase transition should be characterized by a
Hall plateau transition as a function of $\Delta _{LL}$ at a fixed filling
factor. We also study the properties of this crystal state and discuss its
experimental detection.
\end{abstract}

\date{\today }
\maketitle

\section{INTRODUCTION}

Quantum Hall studies of graphene's two dimensional electron system provided
the earliest confirmation of the massless Dirac character of its bands~\cite%
{grapheneexp1,grapeneexp2}. Similar studies have also confirmed the massive
Dirac character of bilayer graphene~bands\cite{mccann,bilayerexp} and are
making progress toward revealing the distinct bands of ABA and ABC trilayers%
\cite{trilayerIQHE}. Few layer graphene systems stacked in different ways
give rise to distinct electronic properties at relevant energy scales which
introduces an entirely new class of two-dimensional electron gas (2DEG)
systems. Recent theoretical work has indicated that except for very weak
fields, multilayer graphene systems\cite{min,pacomultilayer} can be
classified as a new class of 2DEG systems from here on referred to as chiral
2DEGs (C2DEGs)\cite{barlasrevue}. C2DEG models provide an accurate
description of the low-energy properties of few-layer graphene systems with
a variety of different stacking arrangements consistent with symmetries.

Properties of quasiparticle excitations in C2DEGs are determined by their
chirality index $J$. The quasiparticle dispersion is given by $\epsilon
_{J}\sim |p|^{J}$ (where $p$ is momentum quantum number measured about some
special points in the Brillouin zone) and quasiparticles exhibit a Berry
phase of $J\pi $. These properties lead to an unusual Landau quantization.
Due to this, integer quantum Hall effects (IQHEs) in C2DEGs are remarkably
different from that of semiconducting 2DEGs. Examples of this are quantum
Hall effects in single layer, bilayer and trilayer graphene which exhibit
unusual IQHEs described by Dirac continuum models. At low-energies these
systems represent the $J=1$, $J=2$ and $J=3$ instances of the C2DEGs\cite%
{min} model respectively. Another unique property of C2DEGs is the presence
of a zero-energy LL with degenerate LL orbitals $n=0,\cdots ,J-1$. Speaking
loosely, quantum states corresponding to cyclotron orbits with different
radius, which would have different energies in an ordinary two-dimensional
electron gas, are degenerate. This degeneracy is, of course, on top of the
normal Landau level \emph{one-state-per-flux-quantum} degeneracy and the
four-fold degeneracy already present due to spin and valley degrees of
freedom. The zero-energy LL\ is thus $4J$-fold degenerate.

Naively it can be anticipated that Coulomb interactions will lift the $4J$%
-fold degeneracy of the $N=0$ LL by producing spontaneously broken-symmetry
ground states with spin, valley pseudospin and LL orbital pseudospin
polarizations. As a consequence, one would expect quantum Hall plateaus at
all intermediate integer values of the filling factors from $\nu =-2J+1$ to $%
\nu =2J$~\cite{nomura,barlasprl1,FanABCtrilayer}. It turns out that this
expectation only holds for $J\leq 2$ C2DEGs, whereas for $J>2$ insulating
states in the topmost LL appear generically, which leads to a disappearance
of certain Hall plateaus~\cite{barlasprl3} from the above sequence and a
quantization of the Hall conductivity $\sigma _{xy}$ at a value
corresponding to the \textit{adjacent} interaction driven integer quantum
Hall plateau (for example, $\sigma _{xy}=-6e^{2}/h$ when $\nu =-5$). Thus,
the presence of wave functions with different spatial structures appearing
at the same energy has important consequences on the interaction driven QH
states in ABC-trilayer graphene (i.e. C2DEG systems with chirality index $J>2
$). In particular due to the LL orbital degeneracy, spin and valley
polarized non-uniform states are allowed due to charge modulation in the
orbital subspace. As we show these states only appear for $J>2$ C2DEGs, and
are the main focus of this paper.

The existence of these additional plateaus from spontaneously
broken-symmetry ground state has already been confirmed in suspended bilayer
graphene samples and bilayer graphene on SiO$_{2}$/Si substrates\cite%
{octetexperiments} as well as in graphene trilayers\cite%
{tricoucheexperiences}. By applying an electric potential difference $\Delta
_{B}$ (or \textit{bias})\ between the outermost layers (where the low-energy
sites reside) of a graphene bilayer, one can control the population of
electrons in each layer and open a gap $\Delta _{LL}\left( \Delta
_{B},B\right) $ between two adjacent orbital state which is a function of
both $\Delta _{B}$ and the quantizing magnetic field $B.$ The phase diagram
of the C2DEG in the $\nu -\Delta _{LL}$ space has been studied for bilayer
graphene and is very rich. States with spin and/or layer and/or orbital
polarizations are possible\cite%
{barlasprl1,barlasprl2,barlasprl3,coteorbital,cotefouquet}. At $\nu =-1,4,$
the bias drives a series a transitions from an homogeneous state with finite
orbital pseudospin to charge-density-wave states and crystal states where
the orbital pseudospin rotates in space\cite{coteorbital,cotefouquet}.
Interestingly, a similar sequence of transitions has been observed in a thin
film of the helical magnet Fe$_{0.5}$Co$_{0.5}$Si using Lorentz transmission
electron microscopy\cite{xzyu,jhhan,jinpark}. In that system, the phase
transitions are induced by a transverse magnetic field. In the graphene
bilayer, they are induced by $\Delta _{B}$ potential. The Hamiltonian of
both systems are similar. In particular, they contain a Dzyaloshinsky-Moriya
(DM) interaction\cite{dm} which, in the bilayer, is entirely due to exchange
interactions instead of from spin-orbit coupling.

In this paper, we study the quantum Hall ferromagnetic states of the C2DEG
in trilayer graphene's zeroth LL. Assuming the layer and spin degrees of
freedom are inactive~\cite{barlasprl3}, we only consider the orbital
pseudospin at integer filling. This situation occurs at filling factors $\nu
=-5,-4$ and $\nu =4,5$ for finite value of $\left\vert \Delta
_{LL}\right\vert $ or at other filling factors in some specific range of $%
\Delta _{LL}$. The phase diagram consists in a uniform phase which appears
at large absolute values of $\Delta _{LL}$ where electrons occupy the lowest
energy levels and there is no orbital coherence. Upon decreasing the value
of $\left\vert \Delta _{LL}\right\vert $, the uniform state shows an
instability in the pseudospin wave mode at finite wave vectors $\mathbf{q}$
. This instability indicates, in principle, a transition to a unidirectional
charge-density-wave state. We find, however, that this transition is
preempted by a first-order transition to a crystal state with a vortex
pseudospin texture. One remarkable result is that this crystal state in the
trilayer exists at both positive and negative values of $\Delta _{LL}$ in
contrast with the bilayer case where it is found for $\Delta _{LL}<0$ only.
The pseudospin texture of the crystal is more complex than in bilayer
graphene since an electronic state in this system is described not by a CP$%
^{1}$ but by a CP$^{2}$ spinor. This spinor can be decomposed into three
distinct pseudospin vectors. We show that our crystal has a vortex texture
in each of these pseudospins but with different vorticities. We discuss some
properties of the crystal and uniform phases in this article. In particular,
we show that these states can be distinguished by their electromagnetic
absorption spectrum.

Our paper is organized in the following way. In Sec. II, we review the
effective two-band model Hamiltonian for ABC-trilayer graphene. In Sec. III,
we describe the Hartree-Fock and generalized random-phase formalism (GRPA)
used to compute the energy and pseudospin textures of the different phases
as well as their collective excitations. In Sec. IV, we explain our
pseudospin representation for the various phases, define the electric dipole
density and the formalism used to compute the electromagnetic absorption.
The phase diagram of the C2DEG is presented in Sec. V. In Sec. VI we show
that the uniform phase is unstable at a finite wave vector in some range of
bias where the crystal state described in Sec. VII emerges. We discuss the
properties of this crystal state in Sec. VIII and conclude in Sec. IX.
Appendix A lists the values of the Coulomb exchange interactions at zero
wave vector, Appendix B summarizes the Hartree-Fock and GRPA\ equations for
the single and two-particle Green's functions and Appendix C gives the
generators of the group SU(3).

\section{ABC-TRILAYER GRAPHENE AS A C2DEG}

In this section, we describe the low-energy effective Hamiltonian for the
C2DEG in ABC-stacked trilayer graphene. Before we look at the specific
details of ABC-stacked trilayer graphene lets us review the properties of
C2DEGs\cite{barlasrevue}. The low-energy effective Hamiltonian of a chiral
two-dimensional electron gas (C2DEG) can be written\cite{min} as 
\begin{equation}
H_{J}=\xi ^{J}v_{F}p_{c}\left( \frac{p}{p_{c}}\right) ^{J}\left[ \cos \left(
J\theta \right) \sigma _{x}+\sin \left( J\theta \right) \sigma _{y}\right] ,
\label{H1}
\end{equation}%
where $\sigma _{x},\sigma _{y}$ are Pauli matrices, $p$ is the momentum of
the electron, $\theta $ its angle with the $x$ axis and $J$ is the chirality
index. The parameter $p_{c}=\gamma _{1}/v_{0}$ where $\gamma _{1}$ is the
interlayer hopping energy between the two high-energy sites in adjacent
planes and $v_{F}=3c_{0}\gamma _{0}/2\hslash $ is the Fermi velocity with $%
c_{0}$ the separation between carbon atoms in a plane and $\gamma _{0}$ the
intralayer hopping energy between two neighboring carbon atoms. This
Hamiltonian operates in the space of a two-component wave functions $\Psi
_{\pm }$ describing electronic amplitudes on the two low-energy sites $A$
and $B.$ In the valley $K=\left( -2/3,0\right) 2\pi /a,$ $\xi =-1$ and $\Psi
_{-}=\left( \psi \left( A\right) ,\psi \left( B\right) \right) $ whereas in
the valley $K^{\prime }=\left( 2/3,0\right) 2\pi /a,$ $\xi =+1$ and $\Psi
_{+}=\left( \psi \left( B\right) ,\psi \left( A\right) \right) .$ (For an
introduction to the electronic properties of C2DEG's, see Ref. %
\onlinecite{barlasrevue}).

When a transverse magnetic field is applied to a C2DEG, the kinetic energy
of the electrons is quantized into Landau levels with energies 
\begin{equation}
E_{N,\xi }=\mathrm{sgn}\left( N\right) \gamma _{1}\xi \left( \frac{\sqrt{2}%
\hslash v_{F}}{\ell \gamma _{1}}\right) ^{J}\sqrt{\prod\limits_{i=0}^{J-1}%
\left( \left\vert N\right\vert -i\right) },  \label{EJ}
\end{equation}%
where $\mathrm{sgn}$ is the signum function and $\ell =\sqrt{\hslash c/eB}$
is the magnetic length, $N=0,\pm 1,\pm 2,\ldots $ is the LL index The LLs
are 4-fold degenerate when counting valley and spin degrees of freedom. The $%
N=0$ LL is special since, including the valley and spin degrees of freedom,
it is $4J$-fold degenerate. The extra degeneracy comes from the fact that
the eigenspinors in $N=0$ which have the form $\left\{ \left( 
\begin{array}{cc}
0 & h_{n,X}\left( \mathbf{r}\right) 
\end{array}%
\right) ,n=0,1,...,J-1\right\} $ are degenerate (here, $h_{n,X}\left( 
\mathbf{r}\right) $ represents the Landau-gauge wave functions of
conventional 2DEGs given in Eq. (\ref{fh}) below). We refer to the index $n$
as the \textit{orbital} quantum number. The presence of wave functions with
spatial structures that appear at different energies in the ordinary
non-relativistic 2DEG model in the same degenerate manifold can create some
terminological confusion. We will refer to the wave functions $h_{n}$ as
Landau level $n$ orbitals and (as already anticipated) use upper case letter 
$N$ to distinguish levels with different Landau-quantized band energies in
the C2DEG model in a magnetic field.

\begin{figure}[tbph]
\includegraphics[scale=1]{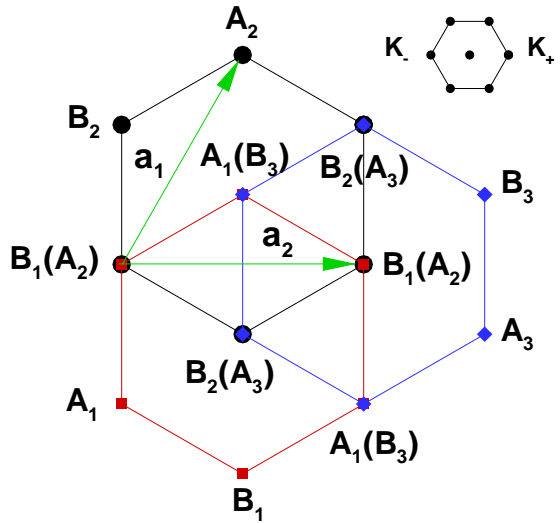}
\caption{(Color online)Lattice structure of an ABC-stacked graphene
trilayer. The two non-equivalent sites of the honeycomb lattice in each
plane are indicated by $A_{m}$ and $B_{m},$ where $m$ is the layer index.
The two basis vectors of the underlying hexagonal Bravais lattice are $%
\mathbf{a}_{1},\mathbf{a}_{2}.$The Brillouin zone of the hexagonal lattice
with the two non-equivalent points $\mathbf{K}_{\pm }$ is drawn in the top
right corner of the figure.}
\label{fig_structuretricouche}
\end{figure}

The lattice structure of an ABC-stacked trilayer graphene (rhombohedral
stacking) is shown in Fig. \ref{fig_structuretricouche}. Each layer has a
honeycomb lattice of carbon atoms. The underlying Bravais lattice is a
triangular lattice with a basis of two atoms denoted by $A_{m}$ and $B_{m}$
where $m$ is the layer index. The triangular lattice constant is $a_{0}=%
\sqrt{3}c_{0}$ where $c_{0}=1.42$ \AA\ is the distance between two
neighboring carbon atoms. The Brillouin zone of the reciprocal lattice has
two non-equivalent $\mathbf{K}$ points that we take as $\mathbf{K}_{\pm
}=\pm \left( 2/3,0\right) 2\pi /a_{0}$ as indicated in the inset of Fig. \ref%
{fig_structuretricouche}. Each adjacent layer pair forms an AB-stacked
bilayer with the upper $B$ sublattice directly on top of the lower $A$
sublattice. The upper $A$ sublattice is above the center of a hexagonal
plaquette of the layer below. Two adjacent layers are separated by a
distance $d=3.35$ \AA .

\begin{figure}[tbph]
\includegraphics[scale=1]{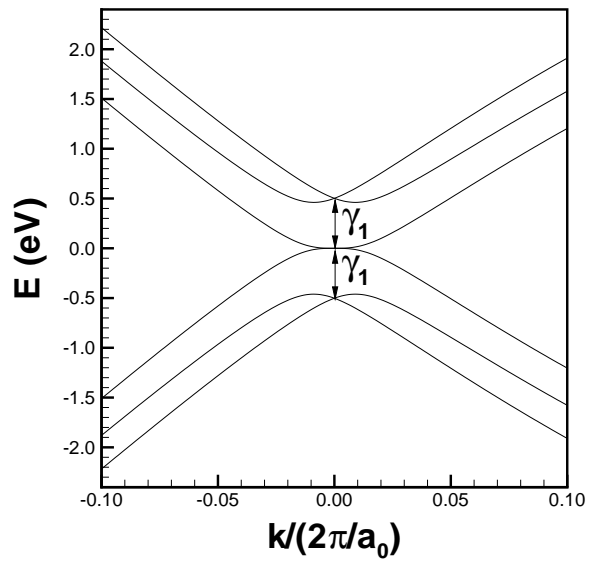}
\caption{ Band structure of ABC-stacked trilayer graphene obtained from the
tight-binding Hamiltonian by keeping the hopping parameters $\protect\gamma %
_{0}$ and $\protect\gamma _{1}$ only.}
\label{fig_bandes}
\end{figure}

The band structure of the ABC-stacked trilayer graphene has been studied in
Refs.\onlinecite{koshino,zhang}. Near the valleys $\mathbf{K}_{\pm }$, it
consists in three valence and three conduction bands as shown in Fig. \ref%
{fig_bandes}. In the simplest model where only the nearest-neighbor
intralayer $\gamma _{0}\approx 3.16$ eV and interlayer hopping $\gamma
_{1}\approx 0.502$ eV are considered, the degenerate bands in the middle of
Fig. \ref{fig_bandes} have a cubic dispersion. For undoped ABC-trilayer
graphene, the valence bands are completely filled and the Fermi level lies
at $E=0$. The high-energy bands are separated by a gap $\gamma _{1}$ from
the low-energy bands as shown in Fig. \ref{fig_bandes}. The low-energy bands
touch at the $\mathbf{K}_{\pm }$ points while the other four bands cross at
the energies $E=\pm \gamma _{1}$ above (below).

To study the low-energy behavior of the electrons, we use an effective
two-band model which results from perturbation theory in $v_{0}^{3}/\gamma
_{1}^{2}$. This effective model can be derived for ABC-stacked trilayer\cite%
{koshino,zhang} starting from a coupled Dirac model, consistent with the
stacking arrangement. In the basis of the low-energy sites $\left(
A_{1},B_{3}\right) $ for the valley $\mathbf{K}_{+}$ and $\left(
B_{3},A_{1}\right) $ for the valley $\mathbf{K}_{-},$ the resulting
Hamiltonian, in the case where a perpendicular magnetic field is applied, is
given by 
\begin{equation}
H_{\xi }^{0}=\left( 
\begin{array}{cc}
\Delta _{\xi }aa^{\dag } & \xi \frac{v_{0}^{3}}{\gamma _{1}^{2}}a^{3} \\ 
\xi \frac{v_{0}^{3}}{\gamma _{1}^{2}}\left( a^{\dag }\right) ^{3} & \Delta
_{-\xi }a^{\dag }a%
\end{array}%
\right) ,  \label{h0}
\end{equation}%
where 
\begin{equation}
\Delta _{\xi }=\xi \frac{\Delta _{B}}{2}-\xi \beta ^{2}\frac{\Delta _{B}}{2},
\label{delta}
\end{equation}%
and $\beta =v_{0}/\gamma _{1}.$In Eq. (\ref{h0}), $a,a^{\dag }$ are the
ladder operators for the Landau levels. The valley index is $\xi =\pm $ and
we have defined $v_{i}=\sqrt{3/2}a_{0}\gamma _{i}/\ell $. In deriving, $%
H_{\xi }^{0},$ we have taken into account a perpendicular electric field
that forces a potential difference $\Delta _{B}$ (or \textit{bias}) between
the outermost layers. In a magnetic field, the low-energy bands of the full
model are replaced by a set of Landau levels with energies, in the absence
of bias, given by Eq. (\ref{EJ}) with $J=2.$These energies are independent
of the guiding-center coordinate $X$ so that each level has the usual
degeneracy $N_{\varphi }=S/2\pi \ell ^{2}$, where $S$ is the area of the
2DEG. The three eigenspinor for the orbital states in $N=0$ and in the
Landau gauge $\mathbf{A}=\left( 0,Bx,0\right) $ are given by 
\begin{equation}
\left( 
\begin{array}{c}
0 \\ 
h_{2,X}\left( \mathbf{r}\right) 
\end{array}%
\right) ,\left( 
\begin{array}{c}
0 \\ 
h_{1,X}\left( \mathbf{r}\right) 
\end{array}%
\right) ,\left( 
\begin{array}{c}
0 \\ 
h_{0,X}\left( \mathbf{r}\right) 
\end{array}%
\right) ,  \label{spinors}
\end{equation}%
where\cite{note} 
\begin{equation}
h_{n,X}\left( \mathbf{r}\right) =\frac{1}{\sqrt{L_{y}}}e^{-iXy/\ell
^{2}}\varphi _{n}\left( x-X\right) ,  \label{fh}
\end{equation}%
are the wave functions in the Landau gauge with $\varphi _{n}\left( x\right) 
$ the wave functions of the one-dimensional harmonic oscillator.

A finite bias lifts this orbital degeneracy. The energies are proportional
to $\Delta _{B}$ and are given as%
\begin{equation}
E_{\xi ,N=0,n}^{0}=-\xi \frac{\Delta _{B}}{2}+n\xi \Delta _{LL},  \label{un}
\end{equation}%
where%
\begin{equation}
\Delta _{LL}=\beta ^{2}\frac{\Delta _{B}}{2}.
\end{equation}%
The correction $\Delta _{LL}$ is small compared to the bias. Indeed, if we
use the values of the tight-binding parameters given in Ref. %
\onlinecite{zhang}, we find $\beta ^{2}=5.\,\allowbreak 49\times 10^{-3}B$
where $B$ is the magnetic field in Tesla. One remarkable aspect of the
energies of the orbital states is that the ordering of the energy levels in $%
N=0$ is different in the two valleys. This property of the energy spectrum
has profound consequences on the phase diagram of the C2DEG in trilayer
graphene, as we will show in this paper.

Other remote inter-layer hopping parameters that have hitherto been
neglected in our analysis can modify the energy spectrum. The hopping term $%
\gamma _{4}$ (which couples the low- and high-energy sites located on
different layers) adds a finite correction $-2n\beta \nu _{4}$ to $E_{\xi
,N=0,n}^{0}$ which is independent of the valley index and bias and scales
linearly with the magnetic field. This correction lifts the degeneracy of
the orbital states even at zero bias. We can include it in $\Delta _{LL}$ by
redefining%
\begin{equation}
\Delta _{LL}=\beta ^{2}\frac{\Delta _{B}}{2}-2\xi \beta \nu _{4}.
\label{deux}
\end{equation}%
In this paper, we take $\Delta _{LL}$ (not $\Delta _{B}$) as the parameter
that we vary to study the phase diagram of the C2DEG. Clearly, $\Delta _{LL}$
can be tuned by changing the bias or the magnetic field. It can have both
positive and negative values. The maximal value of $\left\vert \Delta
_{LL}\right\vert $ must be such that we stay within the limit of validity of
the two-band model. This can be checked by comparing the band structure of
the two-band model (Eqs. (\ref{un}-\ref{deux})) with that given by the full
(six band) model with all hopping terms included. We have done this
comparison and will report it elsewhere\cite{tobe}. Our conclusion is that
there exists a range of bias where the two-band model is well-justified.
This range increases with increasing magnetic field.

\section{ORDER PARAMETERS AND COLLECTIVE EXCITATIONS}

In the present work, we study the phases of the C2DEG when the Fermi level,
filling factor and bias are such that the trilayer can reasonably be
described by a three-level system in the $N=0$ LL with level energies given
by%
\begin{equation}
E_{n}^{0}=n\Delta _{LL}.  \label{ezero}
\end{equation}%
To do this, we must consider the valley and spin degrees of freedom to be
frozen. This can occur, for example, at filling factors $\nu =-5,-4$ or at $%
\nu =4,5$ when the lower levels are fully filled and can be considered as
inert. Also, when Coulomb interaction is included, the Zeeman gap is
exchange-enhanced and, for the filling factors just mentioned, the ground
states were shown to be spin polarized\cite{barlasprl3}. Finally, interlayer
coherence occurs only for very small bias $\left\vert \Delta
_{LL}\right\vert \lesssim 0.001$ $e^{2}/\kappa \ell $ (we checked this
numerically) so that, unless $\Delta _{LL}$ is close to zero, layer
polarization can safely be assumed.

We denote by $\nu _{n}$ the filling factor of the orbital level $n$ and by $%
\widetilde{\nu }$ the filling factor of the three-level system. Our aim is
to study the phase diagram of the C2DEG when $\widetilde{\nu }=1,2$ (since $%
\widetilde{\nu }=3$ is trivial) as $\Delta _{LL}$ is varied.

To study the phase diagram, including both homogeneous and modulated states,
we define the operators%
\begin{eqnarray}
\rho _{n,n^{\prime }}\left( \mathbf{q}\right) &=&\frac{1}{N_{\varphi }}%
\sum_{X,X^{\prime }}e^{-\frac{i}{2}q_{x}\left( X+X^{\prime }\right) } \\
&&\times c_{n,X}^{\dagger }c_{n^{\prime },X^{\prime }}\delta _{X,X^{\prime
}+q_{y}\ell ^{2}},  \notag
\end{eqnarray}%
where $N_{\varphi }=S/2\pi \ell ^{2}$ is the Landau level degeneracy.

The Hartree-Fock Hamiltonian can then be written as

\begin{gather}
H_{HF}=N_{\varphi }E_{n}^{0}\rho _{n,n}\left( 0\right)  \label{HHF2} \\
+N_{\varphi }\overline{\sum_{\mathbf{q}}}H_{n_{1},n_{2},n_{3},n_{4}}\left( 
\mathbf{q}\right) \left\langle \rho _{n_{1},n_{2}}\left( -\mathbf{q}\right)
\right\rangle \rho _{n_{3},n_{4}}\left( \mathbf{q}\right)  \notag \\
-N_{\varphi }\sum_{\mathbf{q}}X_{n_{1},n_{4},n_{3},n_{2}}\left( \mathbf{q}%
\right) \left\langle \rho _{n_{1};n_{2}}\left( -\mathbf{q}\right)
\right\rangle \rho _{n_{3},n_{4}}\left( \mathbf{q}\right) ,  \notag
\end{gather}%
where repeated indices are summed over. In deriving Eq. (\ref{HHF2}), we
have taken into account a neutralizing positive background so that the $%
\mathbf{q}=0$ contribution is absent from the Hartree term. This is
indicated by a bar over the summation.

The Hartree and Fock interactions are defined by 
\begin{eqnarray}
H_{n_{1},n_{2},n_{3},n_{4}}\left( \mathbf{q}\right) &=&\left( \frac{e^{2}}{%
\kappa \ell }\right) \frac{1}{q\ell }K_{n_{1},n_{2}}\left( \mathbf{q}\right)
K_{n_{3},n_{4}}\left( -\mathbf{q}\right) ,  \label{hartree1} \\
X_{n_{1},n_{2},n_{3},n_{4}}\left( \mathbf{q}\right) &=&\int \frac{d\mathbf{p}%
\ell ^{2}}{2\pi }H_{n_{1},n_{2},n_{3},n_{4}}\left( \mathbf{p}\right) e^{i%
\mathbf{q}\times \mathbf{p}\ell ^{2}}.  \label{hartree2}
\end{eqnarray}%
The Coulomb energy $e^{2}/\kappa \ell =56.2\sqrt{B}$ meV with $B$ in Tesla
and $\kappa =1.$ The Fock interactions $X_{n_{1},n_{2},n_{3},n_{4}}\left( 
\mathbf{q}=0\right) $ are listed in Appendix A.

The form factors which appear in $H$ and $X$ are given by

\begin{equation}
K_{n_{1},n_{2}}\left( \mathbf{q}\right) =\left\{ 
\begin{array}{ccc}
F_{n_{1},n_{2}}\left( \mathbf{q}\right) & \mathrm{if} & n_{1}\geq n_{2} \\ 
\left[ F_{n_{2},n_{1}}\left( -\mathbf{q}\right) \right] ^{\ast } & \mathrm{if%
} & n_{1}\leq n_{2}%
\end{array}%
\right. ,  \label{kfactor}
\end{equation}%
with%
\begin{eqnarray}
F_{n,n^{\prime }}\left( \mathbf{q}\right) &=&\sqrt{\frac{n^{\prime }!}{n!}}%
\left( \frac{\left( q_{y}+iq_{x}\right) \ell }{\sqrt{2}}\right)
^{n-n^{\prime }} \\
&&\times e^{-\frac{q^{2}\ell ^{2}}{4}}L_{n^{\prime }}^{n-n^{\prime }}\left( 
\frac{q^{2}\ell ^{2}}{2}\right) .  \notag
\end{eqnarray}%
They capture the character of the different orbital states.

Finally, the Hartree-Fock energy per electron is given by%
\begin{gather}
\frac{E_{HF}}{N_{e}}=\frac{1}{\widetilde{\nu }}E_{n}^{0}\left\langle \rho
_{n,n}\left( 0\right) \right\rangle  \label{HFenergy} \\
+\frac{1}{2\widetilde{\nu }}\overline{\sum_{\mathbf{q}}}%
H_{n_{1},n_{2},n_{3},n_{4}}\left( \mathbf{q}\right) \left\langle \rho
_{n_{1},n_{2}}\left( -\mathbf{q}\right) \right\rangle \left\langle \rho
_{n_{3},n_{4}}\left( \mathbf{q}\right) \right\rangle  \notag \\
-\frac{1}{2\widetilde{\nu }}\sum_{\mathbf{q}}X_{n_{1},n_{4},n_{3},n_{2}}%
\left( \mathbf{q}\right) \left\langle \rho _{n_{1};n_{2}}\left( -\mathbf{q}%
\right) \right\rangle \left\langle \rho _{n_{3},n_{4}}\left( \mathbf{q}%
\right) \right\rangle ,  \notag
\end{gather}%
where $N_{e}$ is the number of electrons in the C2DEG.

The order parameters of the orbital phases are obtained from the
single-particle Matsubara Green's function 
\begin{equation}
G_{n_{1,}n_{2}}\left( X,X^{\prime },\tau \right) =-\left\langle T_{\tau
}c_{n_{1},X}\left( \tau \right) c_{n_{2},X^{\prime }}^{\dagger }\left(
0\right) \right\rangle ,  \label{GFX}
\end{equation}%
where $T_{\tau }$ is the imaginary time ordering operator and $%
c_{n,X}^{\dagger }$ creates an electron in orbital $n$ with guiding-center $%
X.$

If we define the Fourier transform of the single-particle Green's function as

\begin{eqnarray}
G_{n_{1},n_{2}}\left( \mathbf{q,}\tau \right) &=&\frac{1}{N_{\varphi }}%
\sum_{X,X^{\prime }}e^{-\frac{i}{2}q_{x}\left( X+X^{\prime }\right) }
\label{GF} \\
&&\times \delta _{X,X^{\prime }-q_{y}\ell ^{2}}G_{n_{1},n_{2}}\left(
X,X^{\prime },\tau \right) ,  \notag
\end{eqnarray}%
then the order parameters of the coherent phases are simply%
\begin{equation}
\left\langle \rho _{n_{1},n_{2}}\left( \mathbf{q}\right) \right\rangle
=G_{n_{2},n_{1}}\left( \mathbf{q,}\tau =0^{-}\right) .
\end{equation}%
The equation of motion for the Green's function in the Hartree-Fock
approximation is given in Appendix B. This equation leads to the sum rule
(at $T=0$ K)%
\begin{equation}
\sum_{\mathbf{q}}\sum_{n_{2}}\left\vert \left\langle \rho
_{n_{1},n_{2}}\left( \mathbf{q}\right) \right\rangle \right\vert
^{2}=\left\langle \rho _{n_{1},n_{1}}\left( 0\right) \right\rangle .
\label{SR}
\end{equation}%
By definition, we also have%
\begin{equation}
\left\langle \rho _{n,n}\left( 0\right) \right\rangle =\nu _{n}.
\end{equation}

To study the collective excitations, we compute the two-particle Green's
function

\begin{align}
& \chi _{n_{1},n_{2},n_{3},n_{4}}\left( \mathbf{q},\mathbf{q}^{\prime };\tau
\right)  \label{twopart} \\
& =-N_{\varphi }\left\langle T_{\tau }\rho _{n_{1},n_{2}}\left( \mathbf{q,}%
\tau \right) \rho _{n_{3},n_{4}}\left( -\mathbf{q}^{\prime },0\right)
\right\rangle  \notag \\
& +N_{\varphi }\left\langle \rho _{n_{1},n_{2}}\left( \mathbf{q}\right)
\right\rangle \left\langle \rho _{n_{3},n_{4}}\left( -\mathbf{q}^{\prime
}\right) \right\rangle  \notag
\end{align}%
in the generalized random-phase approximation\cite{nozieres} (GRPA). The
resulting set of equations is given in Appendix B. The collective
excitations are given by the poles of the retarded Green's function $\chi
_{n_{1},n_{2},n_{3},n_{4}}^{\left( R\right) }\left( \mathbf{q},\mathbf{q}%
,\omega \right) =\chi _{n_{1},n_{2},n_{3},n_{4}}\left( \mathbf{q},\mathbf{q}%
;i\Omega _{n}\rightarrow \omega +i\delta \right) .$To derive the dispersion
relations, we follow these poles as the wave vector $\mathbf{q}$ is varied
in the Brillouin zone.

\section{DESCRIPTION OF THE ORBITAL-COHERENT PHASES}

\subsection{Pseudospin representation\label{pseudospin}}

An electronic state in our three-state model can be created by the spinor
field%
\begin{equation}
\Phi ^{\dag }\left( \mathbf{r}\right) =\left( 
\begin{array}{c}
\Psi _{0}^{\dag }\left( \mathbf{r}\right) \\ 
\Psi _{1}^{\dag }\left( \mathbf{r}\right) \\ 
\Psi _{2}^{\dag }\left( \mathbf{r}\right)%
\end{array}%
\right) ,  \label{spinor}
\end{equation}%
where $\Psi _{n}^{\dag }\left( \mathbf{r}\right) =\sum_{X}h_{n,X}^{\ast
}\left( \mathbf{r}\right) c_{n,X}^{\dag }.$

Using the eight infinitesimal generators $T_{a}$ (with $a=1,2,...,8$) of
SU(3) (see Appendix C), we can define the eight real fields 
\begin{equation}
\widetilde{F}_{a}\left( \mathbf{r}\right) =\Phi ^{\dag }\left( \mathbf{r}%
\right) T_{a}\Phi \left( \mathbf{r}\right)
\end{equation}%
with the Fourier transforms%
\begin{equation}
\widetilde{F}_{a}\left( \mathbf{q}\right) =\int d\mathbf{r}e^{-i\mathbf{q}%
\cdot \mathbf{r}}\Phi ^{\dag }\left( \mathbf{r}\right) T_{a}\Phi \left( 
\mathbf{r}\right) .
\end{equation}%
With $a=1,$ we get%
\begin{equation}
\widetilde{F}_{1}\left( \mathbf{q}\right) =\frac{1}{2}\left[ N_{\phi
}K_{0,1}\left( -\mathbf{q}\right) \rho _{0,1}\left( \mathbf{q}\right)
+N_{\phi }K_{1,0}\left( -\mathbf{q}\right) \rho _{1,0}\left( \mathbf{q}%
\right) \right] .  \label{fa}
\end{equation}

We describe the three-level system by three pseudospins\cite{elliott} $1/2$.
We associate spin up(down) with level $n=i$($j$) to get the spin $\left(
i,j\right) $ system. We take $\left( i,j\right) =\left( 0,1\right) ;\left(
1,2\right) ;\left( 0,2\right) .$ We suppress the orbital-dependent part of
the form factor and keep only the factor $\beta _{q}=e^{-\frac{q^{2}\ell ^{2}%
}{4}}$ in $K_{i,j}\left( \mathbf{q}\right) $ for all $i,j.$ Eq. (\ref{fa}),
for example, becomes%
\begin{equation}
\widetilde{F}_{1}\left( \mathbf{q}\right) =\beta _{q}\rho _{x}^{\left(
0,1\right) }\left( \mathbf{q}\right) ,  \label{faq}
\end{equation}%
where $\rho _{x}^{\left( i,j\right) }=\frac{1}{2}\left( \rho _{i,j}+\rho
_{j,i}\right) .$ The other components are listed in Appendix C. We also add
to these fields the \textquotedblleft densities\textquotedblright 
\begin{equation}
\rho _{i}\left( \mathbf{q}\right) =\beta _{q}\rho _{i,i}\left( \mathbf{q}%
\right) ,
\end{equation}%
with $i=0,1,2.$

The eight real fields $F_{i}\left( \mathbf{r}\right) $ (the Fourier
transforms of $F_{i}\left( \mathbf{q}\right) $) provide a complete
description of each phase studied in this paper.

\subsection{Dipole density}

The total electronic density is given by 
\begin{equation}
n\left( \mathbf{r}\right) =\sum_{i,j}\Psi _{i}^{\dag }\left( \mathbf{r}%
\right) \Psi _{j}\left( \mathbf{r}\right) .  \label{densityp}
\end{equation}%
Its Fourier transform is 
\begin{equation}
n\left( \mathbf{q}\right) =N_{\varphi }\sum_{i,j=0}^{2}K_{i,j}\left( -%
\mathbf{q}\right) \rho _{i,j}\left( \mathbf{q}\right) .
\end{equation}%
An external electric field, $\mathbf{E}_{ext}\left( \mathbf{r}\right)
=-\nabla \phi _{ext}\left( \mathbf{r}\right) ,$ couples to the density
through a term%
\begin{equation}
H_{ext}=-\frac{e}{S}\sum_{\mathbf{q}}n\left( -\mathbf{q}\right) \phi
_{ext}\left( \mathbf{q}\right) ,
\end{equation}%
in the Hamiltonian where%
\begin{equation}
\phi _{ext}\left( \mathbf{r}\right) =\frac{1}{S}\sum_{\mathbf{q}}\phi
_{ext}\left( \mathbf{q}\right) e^{i\mathbf{q}\cdot \mathbf{r}}.
\end{equation}

Using the definition of the form factors $K_{i,j}\left( \mathbf{q}\right) $
given in Eq. (\ref{kfactor}), the coupling $H_{ext}$ can be written as%
\begin{equation}
H_{ext}=\int d\mathbf{r}\rho _{TOT}\left( \mathbf{r}\right) \phi \left( 
\mathbf{r}\right) -\int d\mathbf{r}\left( \mathbf{d}\left( \mathbf{r}\right)
\cdot \mathbf{E}\left( \mathbf{r}\right) \right) ,
\end{equation}%
where 
\begin{equation}
\rho _{TOT}\left( \mathbf{q}\right) =-e\frac{1}{S}\sum_{i=0}^{2}N_{\varphi
}e^{-q^{2}\ell ^{2}/4}e\rho _{i,i}\left( \mathbf{q}\right) ,  \label{hx}
\end{equation}%
and we can define the dipole operators%
\begin{eqnarray}
d_{x}\left( \mathbf{q}\right) &=&-\gamma \left( \mathbf{q}\right) \left[
\rho _{x}^{\left( 0,1\right) }\left( \mathbf{q}\right) +\alpha \left( 
\mathbf{q}\right) \rho _{x}^{\left( 1,2\right) }\left( \mathbf{q}\right) %
\right] ,  \label{aa_1} \\
d_{y}\left( \mathbf{q}\right) &=&\gamma \left( \mathbf{q}\right) \left[ \rho
_{y}^{\left( 0,1\right) }\left( \mathbf{q}\right) +\alpha \left( \mathbf{q}%
\right) \rho _{y}^{\left( 1,2\right) }\left( \mathbf{q}\right) \right] ,
\label{aa_2}
\end{eqnarray}%
with $\alpha \left( \mathbf{q}\right) =\sqrt{2}\left( 1-q^{2}\ell
^{2}/4\right) $ and $\gamma \left( \mathbf{q}\right) =\sqrt{2}N_{\varphi
}e\ell e^{-q^{2}\ell ^{2}/4}.$

In the phases studied in this paper, $\left\langle \rho _{TOT}\left( \mathbf{%
r}\right) \right\rangle $ is always uniform i.e. $\sum_{i=0}^{2}\left\langle
\rho _{i,i}\left( \mathbf{q}\right) \right\rangle =\widetilde{\nu }\delta _{%
\mathbf{q},0}$ for $\widetilde{\nu }=1,2.$ It follows that we can ignore the
first term in $H_{ext}$ in Eq. (\ref{hx}) and the coupling with the external
electric field is simply%
\begin{equation}
H_{ext}=-\int d\mathbf{r}\left( \mathbf{d}\left( \mathbf{r}\right) \cdot 
\mathbf{E}\left( \mathbf{r}\right) \right) ,
\end{equation}%
where $\mathbf{d}\left( \mathbf{r}\right) $ can be interpreted as a density
of electric dipoles\cite{shizuya1}.

In the absence of Coulomb interaction, the time variation of the total
dipole moment is given by 
\begin{eqnarray}
i\hslash \frac{d}{dt}\mathbf{d}\left( 0\right) &=&-\left[ H_{HF}^{0},\mathbf{%
d}\left( 0\right) \right] \\
&=&i\Delta _{LL}\widehat{\mathbf{z}}\times \mathbf{d}\left( 0\right) , 
\notag
\end{eqnarray}%
where$\allowbreak $%
\begin{equation}
H_{HF}^{0}=N_{\varphi }\left[ \Delta _{LL}\rho _{1,1}\left( 0\right)
+2\Delta _{LL}\rho _{2,2}\left( 0\right) \right] ,
\end{equation}%
so that the dipoles oscillate at the frequency 
\begin{equation}
\omega _{dip}=\Delta _{LL}/\hslash .  \label{dipole3}
\end{equation}%
We can define a dipolar current density by

\begin{equation}
\mathbf{J}^{dip}=\frac{d}{dt}\mathbf{d}\left( 0\right) .
\label{dipolecourant}
\end{equation}

\subsection{Optical absorption}

The total current in the 2DEG is given by%
\begin{equation}
\mathbf{J}=\int d\mathbf{r}\frac{1}{2}\left[ \left( \Psi ^{\dag }\left( 
\mathbf{r}\right) \mathbf{j}\Psi \left( \mathbf{r}\right) \right) +\left( 
\mathbf{j}\Psi \left( \mathbf{r}\right) \right) ^{\dag }\Psi \left( \mathbf{r%
}\right) \right] ,
\end{equation}%
where%
\begin{equation}
\mathbf{j}=-c\left. \frac{\partial H^{0}}{\partial \mathbf{A}^{e}}%
\right\vert _{\mathbf{A}^{e}\rightarrow 0},
\end{equation}%
with $\mathbf{A}^{e}$ the vector potential of the external electromagnetic
field and $H^{0}$ is the Hamiltonian of the two-band model. We find%
\begin{eqnarray}
J_{x} &=&-4\Xi \left[ \rho _{y}^{\left( 0,1\right) }\left( 0\right) +\sqrt{2}%
\rho _{y}^{\left( 1,2\right) }\left( 0\right) \right] ,  \label{courantjx} \\
J_{y} &=&-4\Xi \left[ \rho _{x}^{\left( 0,1\right) }\left( 0\right) +\sqrt{2}%
\rho _{x}^{\left( 1,2\right) }\left( 0\right) \right] ,  \label{courantjy}
\end{eqnarray}%
with the constant%
\begin{equation}
\Xi =N_{\phi }\frac{1}{2\sqrt{2}}\frac{e\ell }{\hslash }\Delta _{LL}.
\end{equation}%
The total current\cite{remark} given by Eqs. (\ref{courantjx}-\ref{courantjy}%
) is nothing but the dipolar current defined in Eq. (\ref{dipolecourant})
above i.e. $\mathbf{J}=\mathbf{J}^{dip}.$

The optical absorption per unit surface from an electromagnetic wave $%
\mathbf{E}=E_{0}\widehat{\mathbf{e}}_{\alpha }e^{i\omega t}$ (with $\alpha
=x,y$ and we define $\overline{\alpha }=x$ if $\alpha =y$ and $\overline{%
\alpha }=y$ if $\alpha =x$) is obtained from the retarded current-current
response function 
\begin{eqnarray}
P_{\alpha }\left( \omega \right) &=&-\frac{1}{\hslash }\Im\left[ \frac{\chi
_{J_{\alpha },J_{\alpha }}^{ret}\left( \omega \right) }{\omega +i\delta }%
\right] E_{0}^{2}  \label{abso} \\
&=&-\frac{2}{h}\left( \frac{eE_{0}\Delta _{LL}}{\hslash }\right) ^{2}Im\left[
\frac{\chi _{\overline{\alpha },\overline{\alpha }}\left( 0,\omega \right) }{%
\omega +i\delta }\right] ,  \notag
\end{eqnarray}%
where the retarded current response function is obtained from the
time-ordered two-particle Green's function%
\begin{equation}
\chi _{J_{\alpha },J_{\beta }}\left( \tau \right) =-\left\langle TJ_{\alpha
}\left( \tau \right) J_{\beta }\left( 0\right) \right\rangle ,
\end{equation}%
with%
\begin{equation}
\chi _{\alpha ,\beta }\left( 0,\tau \right) =-\left\langle T\rho _{\alpha
}\left( 0,\tau \right) \rho _{\beta }\left( 0,0\right) \right\rangle
\end{equation}%
and%
\begin{equation}
\rho _{\alpha }\left( 0,\tau \right) =\rho _{\alpha }^{\left( 0,1\right)
}\left( 0,\tau \right) +\sqrt{2}\rho _{\alpha }^{\left( 1,2\right) }\left(
0,\tau \right) .
\end{equation}

\section{PHASE DIAGRAM OF THE C2DEG}

The ground state in the ABC-trilayer graphene can be classified in terms of
translationally invariant (uniform) states or non-translationally invariant
(non-uniform) states. One important distinction to note is that any
non-uniform state in graphene would otherwise be either spin or valley
density waves. However, for spin and valley polarized $J\geq 2$ C2DEGs,
non-uniform states are generically allowed due to charge modulation in the
orbital subspace. Both states have different experimental signatures:
uniform states exhibit Hall conductivity whereas if the topmost occupied LL
has a crystal-like state or unidirectional charge-density-wave, it will
either be insulating or exhibit anisotropic conductivity. In the case of
crystal-like states discussed here, which will likely be pinned by disorder,
the Hall conductivity $\sigma _{xy}$ will be at a value corresponding to the 
\textit{adjacent} interaction driven integer quantum Hall plateau as we will
discuss.

In our study of the phase diagram, we consider the following states:

\begin{enumerate}
\item A coherent uniform state (CUP). In this state, the only allowed order
parameters are $\left\langle \rho _{n,m}\left( \mathbf{q}=0\right)
\right\rangle .$ The state of each electron is described by the CP$^{2}$
spinor $\left( a_{0},a_{1},a_{2}\right) $ (where $a_{i}$'s are complex
numbers satisfying $\sum_{n}\left\vert a_{n}\right\vert ^{2}=1$) so that an
electron at guiding-center $X$ is in a linear combination of the three
orbital states. This combination is the same for all electrons. The CUP\
ground state is written as 
\begin{equation}
\left\vert \Psi \right\rangle _{CUP}=\prod\limits_{X}\left[
\sum_{n=0}^{2}a_{n}c_{n,X}^{\dag }\right] \left\vert 0\right\rangle
\label{state}
\end{equation}%
which gives 
\begin{equation}
\left\langle \rho _{n,m}\left( \mathbf{q}=0\right) \right\rangle
=a_{n}^{\ast }a_{m}.
\end{equation}%
Due to particle-hole symmetry the ground state for $\widetilde{\nu }=2$ can
be described as a filled level of holes on a vacuum state consisting of the
three levels filled with electrons. The CP$^{2}$ spinor $\left(
a_{0},a_{1},a_{2}\right) $ also applies to an hole state if $c_{nX}^{\dagger
}\rightarrow b_{nX}^{\dagger }$ (where $b_{nX}^{\dagger }$ is a hole
creation operator). The CUP phase is possible at negative bias because the
system can then reduce its kinetic energy by populating the levels $n=1,2$
that are below level $n=0$ in energy. This, however, increases the exchange
energy because the Coulomb exchange terms satisfy $X_{0,0,0,0}\left(
0\right) >X_{1,1,1,1}\left( 0\right) >X_{2,2,2,2}\left( 0\right) $ (see
Appendix A). In consequence, there is an optimal population of the levels
that minimizes the total energy.

\item An incoherent uniform phase (IUP). In this case, the only allowed
order parameters are $\left\langle \rho _{n,n}\left( \mathbf{q}=0\right)
\right\rangle $ and the first ( $\widetilde{\nu }=1$) or first two ( $%
\widetilde{\nu }=2$) lowest-lying orbital states are fully filled so that
all coherences $a_{n}^{\ast }a_{m}$ $\left( n\neq m\right) $ are zero. For $%
\widetilde{\nu }=1$, we have in this limit $a_{0}=1$ when $\Delta _{LL}>0$
and $a_{2}=1$ when $\Delta _{LL}<0.$ (It is just the opposite for the hole
spinor when $\widetilde{\nu }=2.$)

\item A coherent charge-density-wave phase (CCDWP). This state is modulated
in one direction only and the allowed order parameters are $\left\langle
\rho _{n,m}\left( p\mathbf{q}_{0}\right) \right\rangle $ where $p=0,\pm
1,\pm 2,...$ and $\mathbf{q}_{0}$ is the wave vector of the CCDWS. The
ground state is simply obtained by letting $a_{n}\rightarrow a_{n}\left(
X\right) $ in Eq. (\ref{state}).

\item A coherent crystal phase (CCP). In this non-uniform state, all order
parameters $\left\{ \left\langle \rho _{n,m}\left( \mathbf{G}\right)
\right\rangle \right\} $ are allowed where $\left\{ \mathbf{G}\right\} $ are
the reciprocal lattice vectors of the Bravais lattice of the crystal. We
have considered a triangular and a square lattice with one electron per unit
cell.
\end{enumerate}

Note that in all of these states, the "density" $\left\langle \rho \left( 
\mathbf{r}\right) \right\rangle =\sum_{n}\left\langle \rho _{n,n}\left( 
\mathbf{r}\right) \right\rangle $ is a constant in space. The real density $%
\left\langle n\left( \mathbf{r}\right) \right\rangle $ as defined in Eq. (%
\ref{densityp}) is modulated in space in the CCDWP and CCP.

Our Hartree-Fock numerical calculations for the phase diagram of the C2DEG
is shown in Table 1. The CUP and CCDWP\ are never the ground state. The
ground state is an IUP for large value of $\left\vert \Delta
_{LL}\right\vert $ and a CCP\ (with a triangular lattice) in between. The
transitions from the IUP to the crystal state are first order.

\begin{center}
\begin{table}[tbp] \centering%
\begin{tabular}{|l|l|}
\hline
\textbf{Phase} & $\Delta _{LL}$ \\ \hline
IUP in $n=2$ & $\Delta _{LL}<-0.31(-0.34)$ $e^{2}/\kappa \ell $ \\ 
CCP & $-0.31(-0.34)$ $e^{2}/\kappa \ell <\Delta _{LL}<0.09(0.03)$ $%
e^{2}/\kappa \ell $ \\ 
IUP in $n=0$ & $\Delta _{LL}>0.09(0.03)$ $e^{2}/\kappa \ell $ \\ \hline
\end{tabular}%
\caption{Phase diagram including the uniform and crystal phases for $\tilde\nu=1$. The numbers
in parenthesis are for  $\tilde\nu=2$.}\label{Tableau1}%
\end{table}%
\end{center}

We discuss the crystal state in more details in the next section. We point
out here, however, that the presence in the phase diagram of a crystal state
at positive value of $\Delta _{LL}$ is something specific to trilayer
graphene or other C2DEGS with $J > 2$ as discussed in Sec. \ref{absence}
below.

\section{PROPERTIES OF THE UNIFORM PHASES}

Even though the CUP\ is not the ground state in trilayer graphene, we would
like to give a brief description some of its properties. In particular, it
shows an instability to a CCDW\ state that is, in the case of the trilayer,
preempted by the crystal phase. Comparing the energies of the IUP and CUP in
the HFA, we find that the CUP has lower energy than the IUP in the range of
bias $\Delta _{LL}\in \left[ \Delta _{LL}^{(\ast )},0\right] $ where $\Delta
_{LL}^{(\ast )}=-0.29e^{2}/\kappa \ell $ for $\widetilde{\nu }=1$ and $%
\Delta _{LL}^{(\ast )}=-0.24e^{2}/\kappa \ell $ for $\widetilde{\nu }=2.$

\begin{figure}[tbph]
\includegraphics[scale=1]{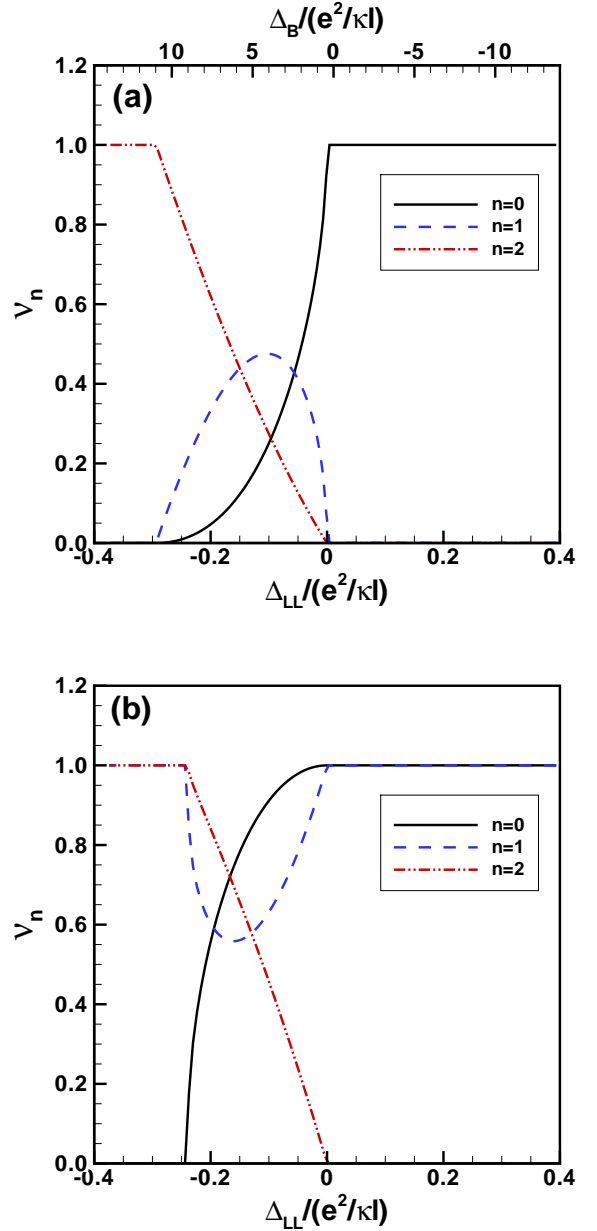}
\caption{(Color online) Occupation of the orbital levels in the incoherent ($%
\Delta _{LL}\geq 0$) and coherent ($\Delta _{LL}\leq 0$) uniform phases for $%
B=10$ T as a function of the bare gap $\Delta _{LL}.$(a) $\widetilde{\protect%
\nu }=1$ and (b) $\widetilde{\protect\nu }=2.$}
\label{fig_populationliquide}
\end{figure}

The ground-state energy per electron\ is an uniform phase is given by 
\begin{align}
& \frac{E_{CUP}}{N_{e}}=\frac{1}{\widetilde{\nu }}\left[ \Delta
_{LL}\left\langle \rho _{1,1}\right\rangle +2\Delta _{LL}\left\langle \rho
_{2,2}\right\rangle \right]   \label{cupsg} \\
& -\frac{1}{2\widetilde{\nu }}\left[ X_{0,0,0,0}\left\langle \rho
_{0,0}\right\rangle ^{2}+X_{1,1,1,1}\left\langle \rho _{1,1}\right\rangle
^{2}+X_{2,2,2,2}\left\langle \rho _{2,2}\right\rangle ^{2}\right]   \notag \\
& -\frac{1}{\widetilde{\nu }}\left[ X_{0,1,1,0}\left\langle \rho
_{0,0}\right\rangle \left\langle \rho _{1,1}\right\rangle
+X_{0,2,2,0}\left\langle \rho _{0,0}\right\rangle \left\langle \rho
_{2,2}\right\rangle \right]   \notag \\
& -\frac{1}{\widetilde{\nu }}\left[ X_{1,2,2,1}\left\langle \rho
_{1,1}\right\rangle \left\langle \rho _{2,2}\right\rangle \right]   \notag \\
& -\frac{1}{\widetilde{\nu }}\left[ X_{0,0,1,1}\left\vert \left\langle \rho
_{0,1}\right\rangle \right\vert ^{2}+X_{0,0,2,2}\left\vert \left\langle \rho
_{0,2}\right\rangle \right\vert ^{2}+X_{1,1,2,2}\left\vert \left\langle \rho
_{1,2}\right\rangle \right\vert ^{2}\right]   \notag \\
& -\frac{1}{\widetilde{\nu }}\left[ X_{0,1,2,1}\left\langle \rho
_{0,1}\right\rangle \left\langle \rho _{2,1}\right\rangle
+X_{1,0,1,2}\left\langle \rho _{1,2}\right\rangle \left\langle \rho
_{1,0}\right\rangle \right]   \notag
\end{align}%
where all $\left\langle \rho _{i,j}\right\rangle \,^{\prime }s$ and the
interactions $X_{i,j,k,l}$ are evaluated at $\mathbf{q}=0$. Note that if we
make a local gauge transformation of the CP$^{2}$ spinor i.e. $a_{i}\left(
X\right) \rightarrow a_{i}e^{i\Lambda \left( X\right) },$ the energy $E_{CUP}
$ is unchanged. This $U(1)$ gauge invariance is necessary to define the CP$%
^{2}$ spinor\cite{goshrajaraman}. Fig. \ref{fig_populationliquide} shows the
occupation of the levels in both phases (IUP\ and CUP) for filling factors $%
\widetilde{\nu }=1,2$.

One possible parametrization of a CP$^{2}$ spinor is given by\cite%
{ezawalivre} 
\begin{equation}
\left( 
\begin{array}{c}
\cos \theta \\ 
e^{i\alpha }\sin \theta \cos \varphi \\ 
e^{i\left( \beta +\alpha \right) }\sin \theta \sin \varphi%
\end{array}%
\right) .
\end{equation}%
The first four lines on the right hand side of Eq. (\ref{cupsg}) are
independent of the angles $\alpha $ and $\beta $ while the two terms in the
last line depend on $\cos \left( \alpha -\beta \right) .$ We have verified
numerically that the ground state energy is minimized when $\alpha =\beta $
and that it is independent of the choice of $\alpha $. The CUP thus has a
broken U(1) symmetry and supports a Goldstone mode. This is confirmed by our
GRPA\ calculation which also shows that the dispersion of this mode is
highly anisotropic. The CUP in a graphene bilayer has similar properties as
reported in Ref. \onlinecite{coteorbital} where the origin of the anisotropy
is discussed.

By contrast, the lowest-energy mode in the IUP is gapped and has an
isotropic dispersion. This gap can be calculated analytically from the GRPA\
equations. We find%
\begin{eqnarray}
&&\omega _{IUP}\left( \mathbf{q}=0\right)  \label{gap} \\
&=&\left\{ 
\begin{tabular}{lll}
$\Delta _{LL},$ & if & $\widetilde{\nu }=1,\Delta _{LL}\geq 0$ \\ 
$-\Delta _{LL}-\frac{15}{64}\sqrt{\frac{\pi }{2}}\left( \frac{e^{2}}{\kappa
\ell }\right) ,$ & if & $\widetilde{\nu }=1,\Delta _{LL}\leq 0$ \\ 
$\Delta _{LL},$ & if & $\widetilde{\nu }=2,\Delta _{LL}\geq 0$ \\ 
$-\Delta _{LL}-\frac{3}{16}\sqrt{\frac{\pi }{2}}\left( \frac{e^{2}}{\kappa
\ell }\right) ,$ & if & $\widetilde{\nu }=2,\Delta _{LL}\leq 0$%
\end{tabular}%
\right. .  \notag
\end{eqnarray}%
The critical bias $\Delta _{LL}^{(\ast )}$ for the transition from the IUP
to the CUP is given by the condition $\omega _{IUP}\left( \mathbf{q}%
=0\right) =0$. The frequency $\omega _{IUP}\left( \mathbf{q}=0\right) $ is
however positive in the region where the IUP is the ground state according
to Table 1. This frequency is measurable in electromagnetic absorption
experiments. We come back to this point in Sec. VIII.

\begin{figure}[tbph]
\includegraphics[scale=1]{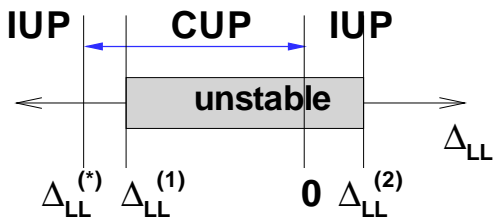}
\caption{(Color online)\ Phase diagram at $\widetilde{\protect\nu }=1$. The
incoherent uniform phase (IUP) occurs for $\Delta _{LL}\geq 0$ and for $%
\Delta _{LL}\leq \Delta _{LL}^{(\ast )}.$ In between these two biases, the
system is in a coherent uniform phase (CUP) in the HFA. In the GRPA, the
collective excitations show that the uniform phases are unstable between $%
\Delta _{LL}^{(1)}$ and $\Delta _{LL}^{(2)}.$}
\label{fig_diaphase}
\end{figure}

\begin{figure}[tbph]
\includegraphics[scale=1]{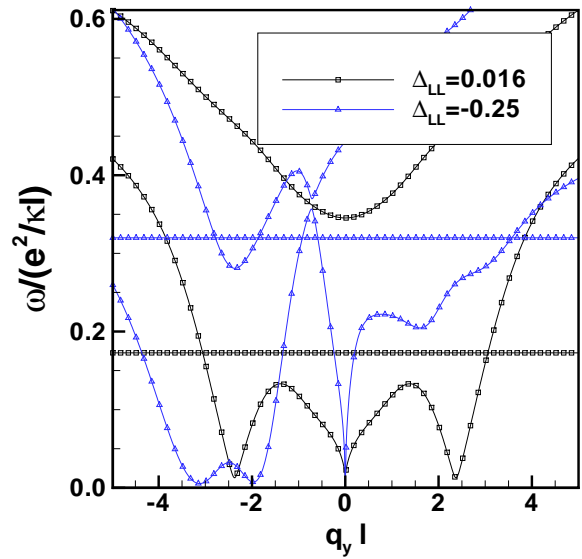}
\caption{(Color online) Dispersion relation of the collective modes in the
uniform phases near the instability points $\Delta _{LL}^{(1)}=-0.25$ $e^{2}/%
\protect\kappa \ell $ (CUP) and $\Delta _{LL}^{(2)}=0.016$ $e^{2}/\protect%
\kappa \ell $ (IUP) defined in Fig.4.}
\label{fig_dispersionliquidenu1}
\end{figure}

The dispersion of the lowest-energy mode in the IUP and CUP becomes unstable
at a finite value of $\mathbf{q}$ in some range of bias $\Delta _{LL}\in %
\left[ \Delta _{LL}^{(1)},\Delta _{LL}^{(2)}\right] .$Fig. \ref{fig_diaphase}
shows the situation for $\widetilde{\nu }=1.$ In this case, the IUP is
unstable for $q\ell \approx 2$ at $\Delta _{LL}^{(2)}=0.016$ $e^{2}/\kappa
\ell $ while the CUP is unstable for $q_{y}\ell \approx -2$ and $q_{y}\ell
\approx -3$ at $\Delta _{LL}^{(1)}=-0.25$ $e^{2}/\kappa \ell .$ The
direction in $\mathbf{q}-$space of the instability is related to the
orientation of the electric dipoles present in the CUP. The instability in $%
\omega \left( \mathbf{q}\right) $ occurs in the direction $\widehat{\mathbf{q%
}}=\widehat{\mathbf{z}}\times \widehat{\mathbf{d}}\left( \mathbf{r}\right) $
as in the bilayer case\cite{coteorbital}. The dispersions at these two
biases are plotted in Fig. \ref{fig_dispersionliquidenu1}. For filling
factor $\widetilde{\nu }=2,$ it is the IUP at \textit{negative} $\Delta
_{LL} $ that becomes unstable for $\Delta _{LL}\geq \Delta _{LL}^{(1)}=-0.27$
$e^{2}/\kappa \ell $ and the instability persists well into the CUP until $%
\Delta _{LL}^{(2)}=-0.036$ $e^{2}/\kappa \ell $. The system is stable for $%
\Delta _{LL}\geq \Delta _{LL}^{(2)}.$

The instability at a finite wave vector suggests that both uniform phases
are unstable towards the formation of some kind of charge-density-wave
state. Some of us discussed this instability in Ref.\onlinecite{barlasprl3}.
For negative value of $\Delta _{LL},$ this is not surprising. The same
situation occurs in bilayer graphene\cite{cotefouquet,coteorbital}. The
instability for \textit{positive} value of $\Delta _{LL}$ is unexpected,
however, and does not occur in bilayer graphene\cite{barlasprl1}.

\section{CRYSTAL STATE WITH\ ORBITAL\ COHERENCE}

In the HFA, we find that the instability towards the CDW state is preempted
by the formation of a coherent crystal phase (CCP) with a triangular
lattice. This electron crystal has one electron ($\widetilde{\nu }=1$) or
one hole ($\widetilde{\nu }=2$) per site so that the lattice constant is $%
a_{0}/\ell =\sqrt{2\pi /\sqrt{3/2}}$ in both cases. The real density $%
n\left( \mathbf{r}\right) $ is modulated in space but not the "density" $%
\left\langle \rho \left( \mathbf{r}\right) \right\rangle
=\sum_{n}\left\langle \rho _{n,n}\left( \mathbf{r}\right) \right\rangle $
which is a constant. At integer filling, this is possible because in our
system, the electrons can be distributed in more than one orbital states.
Every electron in the CCP is in a linear superposition of the three orbital
states as in the CUP\ but also in a superposition of guiding center states $%
X.$ All order parameters $\left\langle \rho _{n,m}\left( \mathbf{G}\right)
\right\rangle $ are finite where $\left\{ \mathbf{G}\right\} $ are the
reciprocal lattice vectors of the crystal. We choose $\left\vert \mathbf{G}%
\right\vert $ big enough in the numerical calculation to insure that the
crystal energy converges to the required accuracy. We show in Fig. \ref%
{fig_crystalnu11} the filling factors $\nu _{n}$ of the orbital states as
well as the cohesive energy of the crystal. This energy is quite large, of
the order of $0.04$ $e^{2}/\kappa \ell \approx 7$ meV$\approx 80$ K at $B=10$
T and for $\kappa =1.$ Hence, this state should be quite robust against
thermal fluctuations and disorder.

\begin{figure}[tbph]
\includegraphics[scale=1]{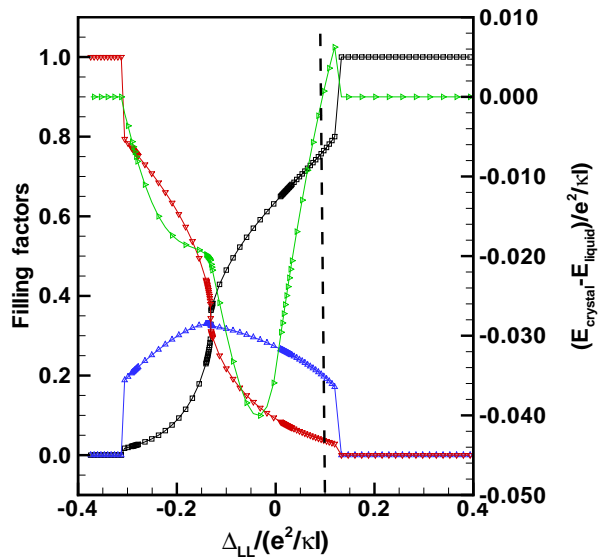}
\caption{(Color online) Filling factors (left $y$ axis) of the orbital
levels $n=0$ (squares), $n=1$ (delta), and $n=2$ (nabla) in the cystal
states for $\widetilde{\protect\nu }=1$ and $B=10$ T. The difference in
energy between the crystal and the uniform states is shown by the green
curve with the right triangles (right $y$ axis). The dashed vertical line
indicates the point where the uniform state is lower in energy than the
crystal state for $\Delta _{LL}$ positive.}
\label{fig_crystalnu11}
\end{figure}

\begin{figure}[tbph]
\includegraphics[scale=0.9]{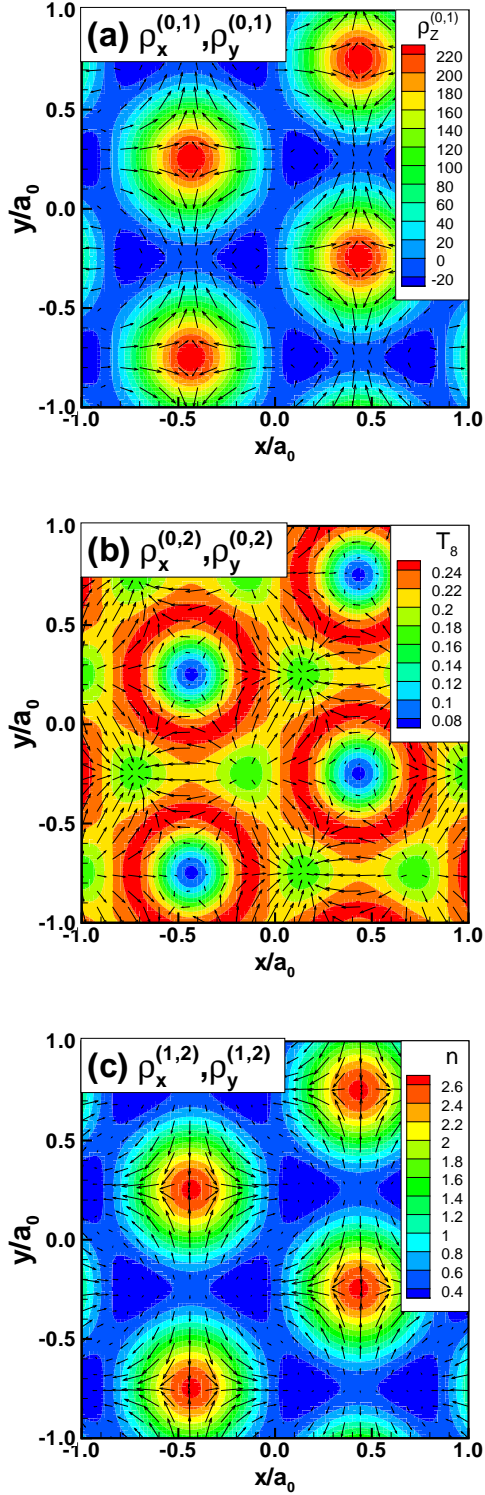}
\caption{(Color online) Pseudospin texture in the (a) $(0,1)$; (b) $(0,2)$;
(c) $(1,2)$ pseudospin systems for the crystal state at $\protect\nu %
=1,\Delta _{LL}=0$ and $B=10$ T. The total electronic density and $%
T_{8}\left( \mathbf{r}\right) $ are shown in the contour plots of (c) and
(d) respectively.}
\label{fig_spintexture}
\end{figure}

We show in Fig. \ref{fig_spintexture} the electronic density $n\left( 
\mathbf{r}\right) $ and the full SU(3)\ pseudospin representation of the
crystal state for $\widetilde{\nu }=1$ at $B=10$ T. We have chosen $\Delta
_{LL}=0$ in this figure, but the textures do not change much as $\Delta _{LL}
$ is varied. We remark that the fields $\widetilde{F}_{a}\left( \mathbf{r}%
\right) $ defined in Eq. (\ref{faq}) represent pseudospin \textit{densities}%
. These are not bounded. In particular, they are not normalized in Fig. \ref%
{fig_spintexture}. We find that for the three pseudospin fields, $%
\left\langle \rho _{z}^{\left( i,j\right) }\left( \mathbf{r}\right)
\right\rangle >>\left\langle \rho _{x}^{\left( i,j\right) }\left( \mathbf{r}%
\right) \right\rangle ,\left\langle \rho _{y}^{\left( i,j\right) }\left( 
\mathbf{r}\right) \right\rangle $ so that the pseudospin are almost
completely polarized in the direction of the $z$ axis or opposite to it. The
in-plane component of the pseudospin vectors (and so the interorbital
coherence) is small. There is however a small clockwise rotation of the
in-plane component of the pseudospins in the $x-y$ plane around each lattice
site. For $\left( i,j\right) =\left( 0,1\right) $ and $\left( 1,2\right) $
the rotation of the pseudospins is of $2\pi $ while for $\left( 0,2\right) $
the pseudospins rotate by $4\pi .$

We conjecture that the interesting pseudospin texture around each lattice
site can be assigned a topological charge. The minimal CP$^{2}$ sigma model
is known to support skyrmion solutions\cite{ezawalivre}. In our case, where
multiple orbitals are considered, the energy of long-wavelength deformations
contains much more terms than the minimal CP$^{2}$ sigma model. In
particular, it involves multiple pseudospin stuffiness. Nevertheless, we
believe that finite energy excitations should fall into different
topological sectors even in this case. More work is needed, however, to
confirm our conjecture.

A more physical quantity to represent graphically is the dipole density
defined in Eqs. (\ref{aa_1},\ref{aa_2}). It is shown in Fig. \ref%
{fig_densitecristal} for filling factors $\widetilde{\nu }=1,2.$ The vector
field of the dipole density has also a vortex structure around each lattice
site. The dipole vectors rotate by $2\pi $ in both cases.

\begin{figure}[tbph]
\includegraphics[scale=0.9]{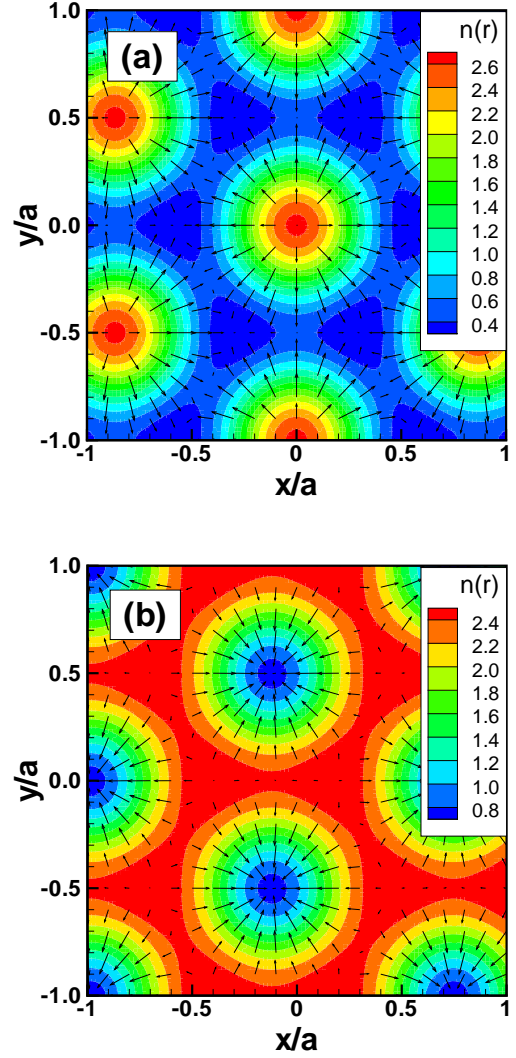}
\caption{(Color online) Electronic density and dipole field in the crystal
state for (a) $\widetilde{\protect\nu }=1$ and (b) $\widetilde{\protect\nu }%
=2.$ Parameters are $\Delta _{LL}=0$ and $B=10$ T.}
\label{fig_densitecristal}
\end{figure}

\subsection{Absence of the crystal state in bilayer graphene\label{absence}}

\begin{figure}[tbph]
\includegraphics[scale=1]{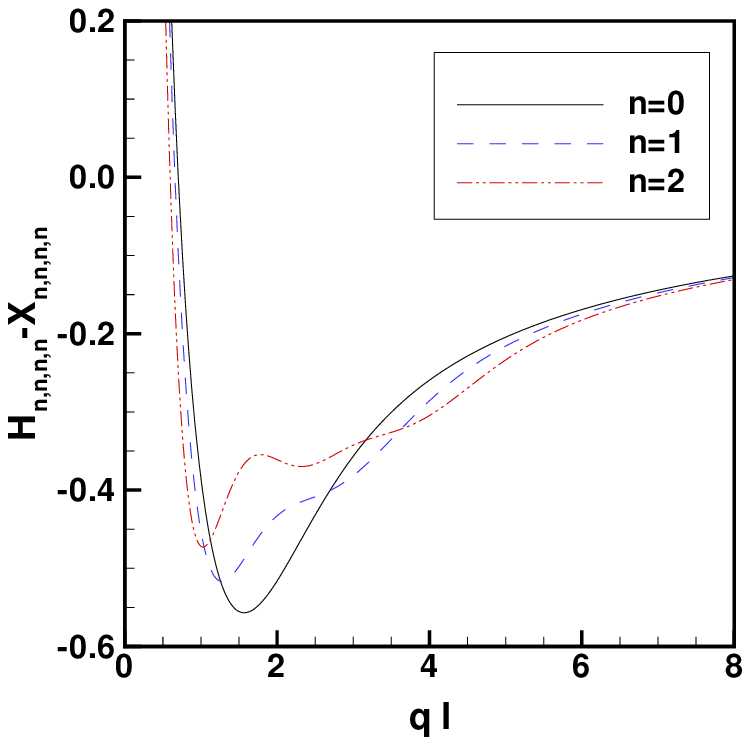}
\caption{(Color online) Effective Hartree-Fock interaction in orbitals $%
n=0,1,2.$}
\label{fig_inter}
\end{figure}

\begin{figure}[tbph]
\includegraphics[scale=0.9]{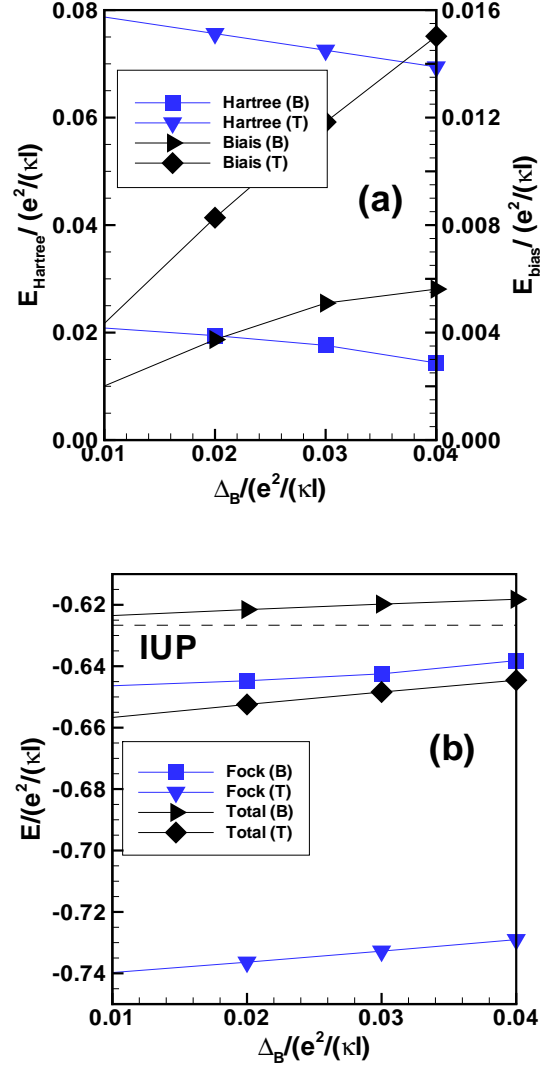}
\caption{(Color online) Contribution of different terms to the total energy
of the crystal state in bilayer (B) and trilayer (T) graphene for $%
\widetilde{\protect\nu }=1$ and $B=10$ T. (a) Hartree and bias energies; (b)
Fock and total energies. The dashed line is the total energy of the
incoherent uniform phase.}
\label{fig_energietricouchebicouche}
\end{figure}

As shown in Fig. \ref{fig_inter}, the effective interactions $%
H_{n_{1},n_{2},n_{3},n_{4}}\left( \mathbf{q}\right)
-X_{n_{1},n_{4},n_{3},n_{2}}\left( \mathbf{q}\right) $ which appears in the
Hartree-Fock energy favor the formation of a CCDWP or CCP state because most
of them take their minimal value at a finite wave vector. Because the
filling factor is an integer, the system must put some of the electrons in
the higher-energy orbital states in order to produce a density modulation.
These modulations increases the Hartree energy of the system and, in the
case where $\Delta _{LL}>0,$ the occupation of the higher-energy levels
increases the bias energy. Nevertheless, the crystal state is favored in the
trilayer because these costs are more than compensated by the gain in
exchange energy as shown in Fig. \ref{fig_energietricouchebicouche}.

In the Bernal-stacked bilayer, the Hartree-Fock equation admits a crystal
solution for $\Delta _{LL}>0$ but its total energy is greater than that of
the IUP (see \ref{fig_energietricouchebicouche} (b)). We believe that the
crystal energy in the bilayer case is not optimal because of the following
reason. The effective interactions are not monotonous as shown in Fig. \ref%
{fig_inter}. At the value $\mathbf{q}_{0}$ where the energy is minimal, the
particular effective interaction $H_{n,n,n,n}\left( \mathbf{q}_{0}\right)
-X_{n,n,n,n}\left( \mathbf{q}_{0}\right) $ is smaller for $n=0.$ At larger
value of $\mathbf{q}$, it is the opposite i.e. the interaction is smaller
for $n=2.$ (The ordering in energy of the interaction $%
H_{n_{1},n_{2},n_{3},n_{4}}\left( \mathbf{q}\right)
-X_{n_{1},n_{4},n_{3},n_{2}}\left( \mathbf{q}\right) $ depends very much on
the indices $n_{1},...,n_{4}$). In a crystal, $\left\langle \rho
_{n_{1},n_{2}}\left( \mathbf{G}\right) \right\rangle $ is non zero for an
infinite set of values of $\mathbf{G}$. These order parameters are
constrained by the sum rules of Eq. (\ref{SR}) and also by the condition $%
\sum_{n}\left\langle \rho _{n,n}\left( \mathbf{r}\right) \right\rangle =%
\widetilde{\nu }$ . In the bilayer case, this condition imposes $%
\left\langle \rho _{0,0}\left( \mathbf{G}\right) \right\rangle
=-\left\langle \rho _{1,1}\left( \mathbf{G}\right) \right\rangle $ for $%
\mathbf{G}\neq 0.$ This severe constraint does not allow the crystal state
to take full advantage of the non-monotonous behavior of the effective
interaction in distributing its weight amongst the order parameters with
different $\mathbf{G}^{\prime }s$. By contrast, the trilayer's constraint $%
\left\langle \rho _{0,0}\left( \mathbf{G}\right) \right\rangle +\left\langle
\rho _{1,1}\left( \mathbf{G}\right) \right\rangle +\left\langle \rho
_{2,2}\left( \mathbf{G}\right) \right\rangle =0$ is much less severe. This
point has been discussed on symmetry grounds in Ref. \onlinecite{barlasprl3}%
. The effective theory describing the CCP to IUP phase transition requires
the presence of third order term which, based on symmetry arguments, vanish
for any two-level system such as bilayer graphene.

\section{PROPERTIES OF THE CRYSTAL STATE}

In this section, we look at some of the properties of the coherent crystal
state in more details.

\subsection{Density of states}

\begin{figure}[tbph]
\includegraphics[scale=1]{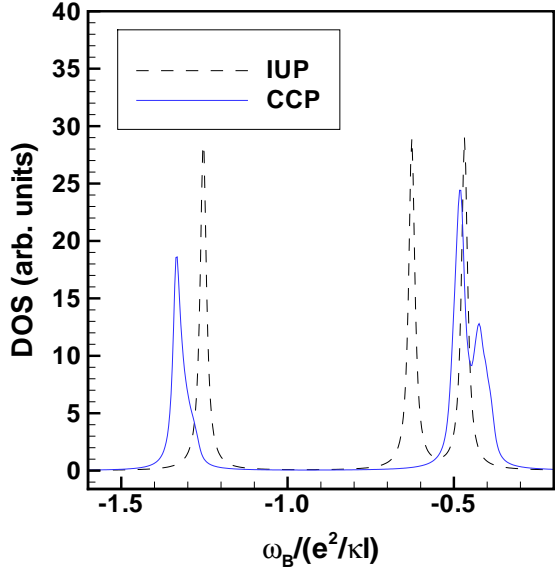}
\caption{(Color\ online)\ Density of states in the incoherent uniform phase
(IUP) and coherent crystal phase (CCP) for $\widetilde{\protect\nu }=1$, $%
\Delta _{LL}=0$ and $B=10$ T.}
\label{fig_dos}
\end{figure}

The density of states in the IUP and CP for $\widetilde{\nu }=1$ and $\Delta
_{LL}=0$,$B=10$ T is shown in Fig. \ref{fig_dos}. In the IUP, the three
peaks correspond to the energies $E_{n}^{0}$ (see Eq. (\ref{ezero})) of the
three levels $n=0,1,2$ renormalized by the exchange interaction. The band
structure of the crystal has three peaks, as expected for a crystal with
filling factor $\widetilde{\nu }=1,$ but slightly displaced in energy and
broadened due to the finite bandwidth of each band in the crystal state. The
lowest-energy band is fully filled while the other two bands are empty. From
this figure, we see that the electron-hole continuum in the crystal state
occurs in the energy range $E\in \left[ 0.6,1.2\right] e^{2}/\kappa \ell $
approximately.

\subsection{Collective excitations, absorption spectrum and quantum Hall
plateaus}

The absorption spectrum for the IUP has one peak at the frequency $\omega
_{IUP}\left( \mathbf{q}=0\right) $ calculated in Eq. (\ref{gap}) which
corresponds to the gap in the first collective mode. The second dispersive
mode shows up in the response functions $\chi _{\rho _{x,y}^{\left(
0,2\right) },\rho _{x,y}^{\left( 0,2\right) }}^{\left( R\right) }$ and since 
$\rho _{x,y}^{\left( 0,2\right) }$ are not part of the dipole definition, it
does not lead to electromagnetic absorption.

The collective mode spectrum in the crystal phase is more complex. Fig. \ref%
{fig_dispersioncristal} shows the dispersion relations for $\widetilde{\nu }%
=1$, $B=10$ T and $\Delta _{LL}=0.$ Only the first low-energy modes that are
below the electron-hole continuum are shown. The first, gapless, mode is the
magnetophonon mode. It has the typical $\omega \sim q^{3/2}$ dispersion
associated with the magnetophonon mode of a Wigner crystal\cite{bonsall}.
The other modes are gapped and correspond to more local deformations of the
density. All modes are accompanied by fluctuations of the pseudospins.

\begin{figure}[tbph]
\includegraphics[scale=1]{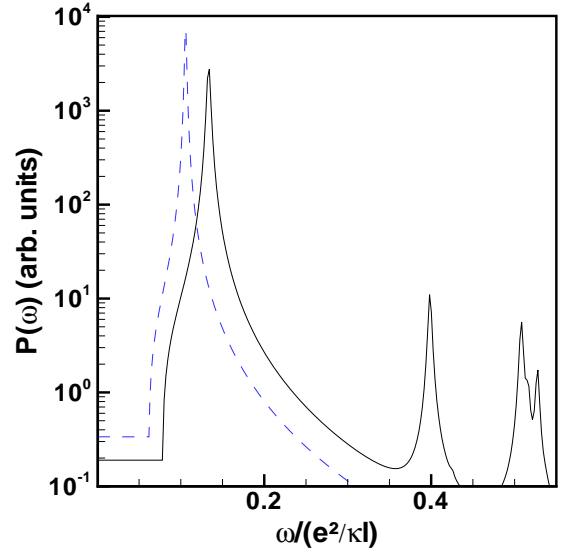}
\caption{(Color online) Comparison of the absorption in the incoherent
uniform phase at $\Delta _{LL}=0.1e^{2}/\protect\kappa \ell $ (dashed blue
line) and in the coherent crystal phase (full black line) at $\Delta
_{LL}=0.09e^{2}/\protect\kappa \ell $. We choose $\widetilde{\protect\nu }=1$
and $B=10$ T in both cases. }
\label{fig_absorption}
\end{figure}

We can get an idea of the nature of the mode at $\mathbf{q}=0$ by computing
the response functions $\chi _{\rho _{\alpha }^{\left( i,j\right) },\rho
_{\alpha }^{\left( i,j\right) }}^{\left( R\right) }\left( \mathbf{q}=0,%
\mathbf{q}=0,\omega \right) $ with $\alpha =x,y$ and $\left( i,j\right)
=\left( 0,1\right) ,\left( 1,2\right) ,\left( 0,2\right) $ from the
pseudospins defined in Sec. IV. We find that the first gapped mode appears
as a pole of $\chi _{\rho _{x}^{\left( 0,1\right) },\rho _{x}^{\left(
0,1\right) }}^{\left( R\right) }$ and $\chi _{\rho _{x}^{\left( 1,2\right)
},\rho _{x}^{\left( 1,2\right) }}^{\left( R\right) }$ but not of $\chi
_{\rho _{x}^{\left( 0,2\right) },\rho _{x}^{\left( 0,2\right) }}^{\left(
R\right) }$ while it is just the opposite for the second gapped mode. We can
thus expect that the first gapped mode will be active in the absorption
while the second gapped mode will not. This is confirmed by a direct
calculation of the absorption spectrum $P_{x}\left( \omega \right) $ as
shown in Fig. \ref{fig_absorption} for the crystal state at $\Delta
_{LL}=0.09e^{2}/\kappa \ell $. Fig. \ref{fig_absorption} also shows the
change in the absorption spectrum when $\Delta _{LL}$ is increased from $%
\Delta _{LL}=0.09e^{2}/\kappa \ell $, in the crystal phase, to $\Delta
_{LL}=0.1e^{2}/\kappa \ell $ where the C2DEG\ has transited to the
incoherent uniform phase. Such a change should be observable experimentally.

\begin{figure}[tbph]
\includegraphics[scale=1]{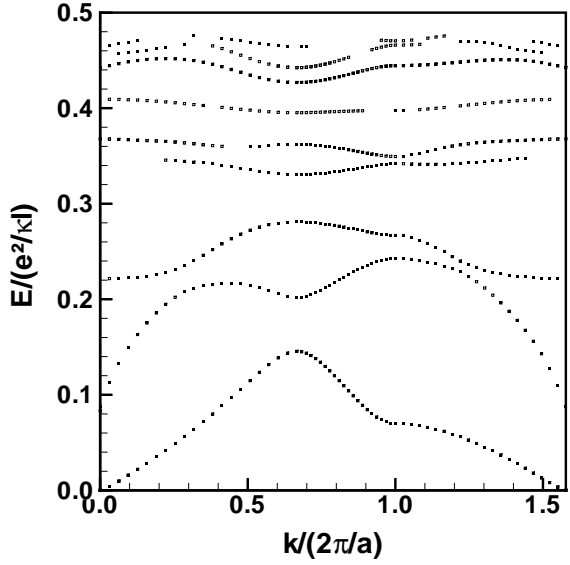}
\caption{Dispersion relation of the collective modes of the crystal phase at 
$\widetilde{\protect\nu }=1,\Delta _{LL}=0$ and $B=10$ T.}
\label{fig_dispersioncristal}
\end{figure}

The Coulomb energy $e^{2}/\kappa \ell =56.2\sqrt{B}$ meV with $B$ in Tesla
and $\kappa =1.$At $B=10$ T, this gives $e^{2}/\kappa \ell =4.3\times
10^{13} $ Hz. The frequency of the first gapped mode is at the upper limit
of the microwave spectrum. It can be pushed down, however, by increasing the
dielectric constant $\kappa $ of the substrate. Note that the crystal state
would likely be pinned by disorder. In this case, the magnetophonon mode
would be gapped and there would be a corresponding absorption at the pinning
frequency corresponding to this gap. The pinning frequency would then depend
on the level of disorder in the system.

The transition from the IUP\ to CP should also show up in the Hall
conductivity. As we just said, the crystal state would be pinned by disorder
leading to insulating behavior of the electrons in the uppermost Landau
level. When $\widetilde{\nu }=1,2$, this means that the quantized Hall
conductivity should have a value corresponding to the \textit{adjacent}
interaction driven integer quantum Hall plateau. For example, if the ground
state stays crystalline in a small region around filling factor $\nu =-5$
(corresponding to $\widetilde{\nu }=1$), there should be a dip in the Hall
plateau from $\sigma _{xy}=-5e^{2}/h$ to $\sigma _{xy}=-6e^{2}/h$ in this
region.

\section{CONCLUSION}

We have studied the phase diagram of the two-dimensional electron gas in an
ABC-stacked graphene trilayer in Landau level $N=0.$ Our analysis is
restricted to the integer filling factors such as $\nu =-5,-4$ and $\nu =4,5$
where spin and layer pseudospin degrees of freedom can be considered as
frozen. We used a three-level system consisting of three Landau orbitals $%
n=0,1,2$ which are separated by an energy gap $\Delta _{LL}$ (this gap is
related to an applied potential bias between the outermost layers) and
considered Coulomb interaction in the Hartree-Fock approximation.

We calculated the energies of different uniform and non-uniform phases of
the C2DEG as $\Delta _{LL}$ is varied from negative to positive values at
integer filling factors $\widetilde{\nu }=1,2$ of the three-level system.
Our results show a phase diagram for the ABC-stacked trilayer graphene that
is very different from its bilayer cousin. The ground state in the trilayer
is an incoherent uniform phase at large value of $\left\vert \Delta
_{LL}\right\vert $ and a coherent crystal phase in between. In the
incoherent uniform phase, electrons fully occupy the lowest-energy orbitals.
In the coherent crystal phase, there is one electron per unit cell of a
triangular crystal lattice and this electron is in a linear combination of
the three orbital states. Around each lattice site of the crystal, we find
an intricate orbital pseudospin texture with a finite density of electric
dipoles in the plane of the layers. The fact that the crystal state also
occurs at positive value of $\Delta _{LL}$ is specific to trilayer graphene
or, more generally, to C2DEG with chirality index $J>2.$

We studied some transport and optical properties of the two phases involved
in the phase diagram. The crystal has a phonon mode that, when disorder is
included in the calculation, should be gapped at $\mathbf{q}=0.$ The
corresponding pinning mode should then be observable in microwave absorption
experiment. The crystal has also higher-energy modes which are active in
electromagnetic absorption. By contrast, only one mode should show up in the
absorption spectrum of the incoherent uniform phase. It should thus be
possible to localize the transition between these two states experimentally.

Another experimental signature of the crystal state is that, if disorder
localizes all electrons in the uppermost Landau level, then a measurement of
the Hall conductivity should show a quantization at a value corresponding to
the \textit{adjacent} interaction driven integer quantum Hall plateau.

\begin{acknowledgments}
R. C\^{o}t\'{e} was supported by a grant from the Natural Sciences and
Engineering Research Council of Canada (NSERC). Computational resources were
provided by Compute Canada and Calcul Qu\'{e}bec.
\end{acknowledgments}

\appendix{}

\section{FOCK INTERACTIONS IN THE UNIFORM PHASES}

In the uniform phases, we need to evaluate the Hartree and Fock interactions
for $\mathbf{q}=0.$ The Hartree interactions are zero and the only nonzero
Fock interactions are (in units of $\sqrt{\pi /2}e^{2}/\kappa \ell $):%
\begin{equation}
X_{0,0,0,0}\left( 0\right) =1,X_{1,1,1,1}\left( 0\right) =\frac{3}{4},
\end{equation}%
\begin{equation}
X_{2,2,2,2}\left( 0\right) =\frac{41}{64},
\end{equation}%
\begin{equation}
X_{0,0,2,2}\left( 0\right) =X_{2,2,0,0}\left( 0\right) =\frac{3}{8},
\end{equation}%
\begin{equation}
X_{2,0,0,2}\left( 0\right) =X_{0,2,2,0}\left( 0\right) =\frac{3}{8},
\end{equation}%
\begin{equation}
X_{2,2,1,1}\left( 0\right) =X_{1,1,2,2}\left( 0\right) =\frac{7}{16},
\end{equation}%
\begin{equation}
X_{1,2,2,1}\left( 0\right) =X_{2,1,1,2}\left( 0\right) =\frac{7}{16},
\end{equation}%
\begin{equation}
X_{0,0,1,1}\left( 0\right) =X_{1,1,0,0}\left( 0\right) =\frac{1}{2},
\end{equation}%
\begin{equation}
X_{1,0,0,1}\left( 0\right) =X_{0,1,1,0}\left( 0\right) =\frac{1}{2},
\end{equation}%
\begin{equation}
X_{0,1,2,1}\left( 0\right) =X_{1,0,1,2}\left( 0\right) =\frac{1}{\sqrt{32}},
\end{equation}%
\begin{equation}
X_{1,2,1,0}\left( 0\right) =X_{2,1,0,1}\left( 0\right) =\frac{1}{\sqrt{32}}.
\end{equation}

\section{HARTREE-FOCK\ AND GRPA\ EQUATIONS}

The equation of motion for the Green's function in the Matsubara formalism
and in the Hartree-Fock approximation is given by 
\begin{gather}
\left( i\omega _{n}+\mu /\hslash -E_{n_{1}}^{0}\right) G_{n_{1},n_{2}}\left( 
\mathbf{q},i\omega _{n}\right)  \label{EHF} \\
-\sum_{\mathbf{q}^{\prime }}\gamma _{\mathbf{q},\mathbf{q}^{\prime
}}U_{n_{1},n_{3}}^{\left( H\right) }\left( \mathbf{q-q}^{\prime }\right)
G_{n_{3},n_{2}}\left( \mathbf{q},i\omega _{n}\right)  \notag \\
+\sum_{\mathbf{q}^{\prime }}\gamma _{\mathbf{q},\mathbf{q}^{\prime
}}U_{n_{1},n_{3}}^{\left( F\right) }\left( \mathbf{q-q}^{\prime }\right)
G_{n_{3},n_{2}}\left( \mathbf{q},i\omega _{n}\right)  \notag \\
=\delta _{n_{1},n_{2}}\delta _{\mathbf{q},0},  \notag
\end{gather}%
where 
\begin{equation}
\gamma _{\mathbf{q},\mathbf{q}^{\prime }}=e^{-i\mathbf{q}\times \mathbf{q}%
^{\prime }\ell ^{2}/2}
\end{equation}%
and%
\begin{eqnarray}
U_{n_{3},n_{4}}^{H}\left( \mathbf{q}\right)
&=&H_{n_{1},n_{2},n_{3},n_{4}}\left( -\mathbf{q}\right) \left\langle \rho
_{n_{1},n_{2}}\left( \mathbf{q}\right) \right\rangle , \\
U_{n_{3},n_{4}}^{F}\left( \mathbf{q}\right)
&=&X_{n_{1},n_{4},n_{3},n_{2}}\left( -\mathbf{q}\right) \left\langle \rho
_{n_{1},n_{2}}\left( \mathbf{q}\right) \right\rangle .
\end{eqnarray}

The self-consistent Eq. (\ref{EHF}) can be put in a matrix form by defining
super-indices. It must then be solved numerically in an iterative way in
order to get the order parameters in the different orbital phases.

In the GRPA, $\chi _{n_{1},n_{2},n_{3},n_{4}}\left( \mathbf{q},\mathbf{q}%
^{\prime };\tau \right) $ is the solution of the equation

\begin{eqnarray}
&&\chi _{n_{1},n_{2},n_{3},n_{4}}\left( \mathbf{q},\mathbf{q}^{\prime
};i\Omega _{n}\right)  \label{grpa1} \\
&=&\chi _{n_{1},n_{2},n_{3},n_{4}}^{\left( 0\right) }\left( \mathbf{q},%
\mathbf{q}^{\prime };i\Omega _{n}\right)  \notag \\
&&+\frac{1}{\hslash }\sum_{\mathbf{q}^{\prime \prime }}\chi
_{n_{1},n_{2},n_{5},n_{6}}^{\left( 0\right) }\left( \mathbf{q},\mathbf{q}%
^{\prime \prime };i\Omega _{n}\right)  \notag \\
&&\times H_{n_{5},n_{6},n_{7},n_{8}}\left( \mathbf{q}^{\prime \prime
}\right) \chi _{n_{7},n_{8},n_{3},n_{4}}\left( \mathbf{q}^{\prime \prime },%
\mathbf{q}^{\prime };i\Omega _{n}\right)  \notag \\
&&-\frac{1}{\hslash }\sum_{\mathbf{q}^{\prime \prime }}\chi
_{n_{1},n_{2},n_{5},n_{6}}^{\left( 0\right) }\left( \mathbf{q},\mathbf{q}%
^{\prime \prime };i\Omega _{n}\right)  \notag \\
&&\times X_{n_{5},n_{8},n_{7},n_{6}}\left( \mathbf{q}^{\prime \prime
}\right) \chi _{n_{7},n_{8},n_{3},n_{4}}\left( \mathbf{q}^{\prime \prime },%
\mathbf{q}^{\prime };i\Omega _{n}\right) ,  \notag
\end{eqnarray}%
where $\Omega _{n}$ is a bosonic Matsubura frequency and the Hartree-Fock
two-particle Green's function $\chi _{n_{1},n_{2},n_{3},n_{4}}^{\left(
0\right) }\left( \mathbf{q},\mathbf{q}^{\prime };i\Omega _{n}\right) $ is
given by

\begin{eqnarray}
&&\left[ i\hslash \Omega _{n}-\left( E_{n_{2}}^{0}-E_{n_{1}}^{0}\right) %
\right] \chi _{n_{1},n_{2},n_{3},n_{4}}^{\left( 0\right) }\left( \mathbf{q},%
\mathbf{q}^{\prime },\Omega _{n}\right)  \label{grpa2} \\
&=&\hslash \gamma _{\mathbf{q},\mathbf{q}^{\prime }}^{\ast }\left\langle
\rho _{n_{1},n_{4}}\left( \mathbf{q-q}^{\prime }\right) \right\rangle \delta
_{n_{2},n_{3}}  \notag \\
&&-\hslash \gamma _{\mathbf{q},\mathbf{q}^{\prime }}\left\langle \rho
_{n_{3},n_{2}}\left( \mathbf{q-q}^{\prime }\right) \right\rangle \delta
_{n_{1},n_{4}}  \notag \\
&&-\overline{\sum_{\mathbf{q}^{\prime \prime }}}\gamma _{\mathbf{q},\mathbf{q%
}^{\prime \prime }}^{\ast }U_{m,n_{1}}^{\left( H\right) }\left( \mathbf{q-q}%
^{\prime \prime }\right) \chi _{m,n_{2},n_{3},n_{4}}^{\left( 0\right)
}\left( \mathbf{q}^{\prime \prime },\mathbf{q}^{\prime },\Omega _{n}\right) 
\notag \\
&&+\overline{\sum_{\mathbf{q}^{\prime \prime }}}\gamma _{\mathbf{q},\mathbf{q%
}^{\prime \prime }}U_{n_{2},m}^{\left( H\right) }\left( \mathbf{q-q}^{\prime
\prime }\right) \chi _{n_{1},m,n_{3},n_{4}}^{\left( 0\right) }\left( \mathbf{%
q}^{\prime \prime },\mathbf{q}^{\prime },\Omega _{n}\right)  \notag \\
&&+\sum_{\mathbf{q}^{\prime \prime }}\gamma _{\mathbf{q},\mathbf{q}^{\prime
\prime }}^{\ast }U_{m,n_{1}}^{\left( F\right) }\left( \mathbf{q-q}^{\prime
\prime }\right) \chi _{m,n_{2},n_{3},n_{4}}^{\left( 0\right) }\left( \mathbf{%
q}^{\prime \prime },\mathbf{q}^{\prime },\Omega _{n}\right)  \notag \\
&&-\sum_{\mathbf{q}^{\prime \prime }}\gamma _{\mathbf{q},\mathbf{q}^{\prime
\prime }}U_{n_{2},m}^{\left( F\right) }\left( \mathbf{q-q}^{\prime \prime
}\right) \chi _{n_{1},m,n_{3},n_{4}}^{\left( 0\right) }\left( \mathbf{q}%
^{\prime \prime },\mathbf{q}^{\prime },\Omega _{n}\right) .  \notag
\end{eqnarray}%
Note that the response functions depend only on the order parameters $%
\left\langle \rho _{n,m}\left( \mathbf{q}\right) \right\rangle $ computed in
the HFA. Eqs. (\ref{grpa1},\ref{grpa2}) can be solved numerically by writing
them in a matrix form defining super-indices.

\section{INFINITESIMAL GENERATORS OF SU(3)\qquad}

Our system has a SU(3) representation. The infinitesimal generators of this
representation are given by the traceless Hermitian matrices%
\begin{equation}
T_{a}=\frac{\lambda _{a}}{2},
\end{equation}%
where, in the basis $\left( 0,1,2\right) $:%
\begin{equation}
\lambda _{1}=\left( 
\begin{array}{ccc}
0 & 1 & 0 \\ 
1 & 0 & 0 \\ 
0 & 0 & 0%
\end{array}%
\right) ,\lambda _{2}=\left( 
\begin{array}{ccc}
0 & -i & 0 \\ 
i & 0 & 0 \\ 
0 & 0 & 0%
\end{array}%
\right) ,
\end{equation}%
\begin{equation}
\lambda _{3}=\left( 
\begin{array}{ccc}
1 & 0 & 0 \\ 
0 & -1 & 0 \\ 
0 & 0 & 0%
\end{array}%
\right) ,\lambda _{4}=\left( 
\begin{array}{ccc}
0 & 0 & 1 \\ 
0 & 0 & 0 \\ 
1 & 0 & 0%
\end{array}%
\right) ,
\end{equation}%
\begin{equation}
\lambda _{5}=\left( 
\begin{array}{ccc}
0 & 0 & -i \\ 
0 & 0 & 0 \\ 
i & 0 & 0%
\end{array}%
\right) ,\lambda _{6}=\left( 
\begin{array}{ccc}
0 & 0 & 0 \\ 
0 & 0 & 1 \\ 
0 & 1 & 0%
\end{array}%
\right) ,
\end{equation}%
and%
\begin{equation}
\lambda _{7}=\left( 
\begin{array}{ccc}
0 & 0 & 0 \\ 
0 & 0 & -i \\ 
0 & i & 0%
\end{array}%
\right) ,\lambda _{8}=\frac{1}{\sqrt{3}}\left( 
\begin{array}{ccc}
1 & 0 & 0 \\ 
0 & 1 & 0 \\ 
0 & 0 & -2%
\end{array}%
\right) .
\end{equation}

With these generators, we can define eight real fields $\widetilde{F}%
_{a}\left( \mathbf{r}\right) =\Phi ^{\dag }\left( \mathbf{r}\right)
T_{a}\Phi \left( \mathbf{r}\right) $ with the Fourier transform%
\begin{equation}
\widetilde{F}_{a}\left( \mathbf{q}\right) =\int d\mathbf{r}e^{-i\mathbf{q}%
\cdot \mathbf{r}}\Phi ^{\dag }\left( \mathbf{r}\right) T_{a}\Phi \left( 
\mathbf{r}\right) .
\end{equation}%
These fields are related to the pseudospin fields by the relations:

\begin{eqnarray}
F_{1}\left( \mathbf{q}\right) &=&\beta _{q}\rho _{x}^{\left( 0,1\right)
}\left( \mathbf{q}\right) ,F_{2}\left( \mathbf{q}\right) =\beta _{q}\rho
_{y}^{\left( 0,1\right) }\left( \mathbf{q}\right) , \\
F_{3}\left( \mathbf{q}\right) &=&\beta _{q}\rho _{z}^{\left( 0,1\right)
}\left( \mathbf{q}\right) ,
\end{eqnarray}%
\begin{equation}
F_{4}\left( \mathbf{q}\right) =\beta _{q}\rho _{x}^{\left( 0,2\right)
}\left( \mathbf{q}\right) ,F_{5}\left( \mathbf{q}\right) =\beta _{q}\rho
_{y}^{\left( 0,2\right) }\left( \mathbf{q}\right) ,
\end{equation}%
\begin{equation}
F_{6}\left( \mathbf{q}\right) =\beta _{q}\rho _{x}^{\left( 1,2\right)
}\left( \mathbf{q}\right) ,F_{7}\left( \mathbf{q}\right) =\beta _{q}\rho
_{y}^{\left( 1,2\right) }\left( \mathbf{q}\right) ,
\end{equation}%
and%
\begin{equation}
F_{8}\left( \mathbf{q}\right) =\frac{1}{2\sqrt{3}}\beta _{q}\left( \rho
_{0,0}\left( \mathbf{q}\right) +\rho _{1,1}\left( \mathbf{q}\right) -2\rho
_{2,2}\left( \mathbf{q}\right) \right) ,
\end{equation}%
where we have defined 
\begin{eqnarray}
\rho _{x}^{\left( i,j\right) } &=&\frac{1}{2}\left( \rho _{i,j}+\rho
_{j,i}\right) , \\
\rho _{y}^{\left( i,j\right) } &=&\frac{1}{2i}\left( \rho _{i,j}-\rho
_{j,i}\right) , \\
\rho _{z}^{\left( i,j\right) } &=&\frac{1}{2}\left( \rho _{i,i}-\rho
_{j,j}\right) .
\end{eqnarray}


\begin{thebibliography}{99}
\bibitem{note} We remark that because of our particular definition of the
ladder operators, we have $a^{\dag }h_{n,X}\left( \mathbf{r}\right) =i\sqrt{%
n+1}h_{n+1,X}\left( \mathbf{r}\right) $ and $ah_{n,X}\left( \mathbf{r}%
\right) =-i\sqrt{n}h_{n-1,X}\left( \mathbf{r}\right) .$
\bibitem{remark} For $\mathbf{q}\neq 0,$ the total current also contains a
contribution that can be written as the curl of a magnetization density.
\end{thebibliography}
\end{document}